\documentclass[twocolumn]{aastex63}

\usepackage{comment}
\usepackage{xspace}
\usepackage{graphicx}
\usepackage{rotating}
\usepackage{hyperref}
\usepackage{amsmath}
\usepackage{amssymb}
\usepackage{chemformula}
\usepackage{color, colortbl}
\usepackage{subfigure}
\newcommand{\lya}{Lyman-\ensuremath{\alpha}\xspace}

\newcommand\Atmos{\texttt{Atmos}\xspace}
\newcommand{\ce}[1]{\ch{#1}}
\newcommand{\gcell}{\cellcolor{gray!25}}

\newcommand{\exotransmit}{\texttt{Exo-Transmit}\xspace}

\definecolor{Gray}{gray}{0.9}

\newcommand{\revision}[2]{\textcolor{red}{\sout{#1}}\textcolor{blue}{\textbf{#2}}}

\renewcommand{\revision}[2]{#2}

\submitjournal{ApJ}

\shortauthors{Teal et al.}
\graphicspath{{./}{figures/}}

\begin{document}

\title{Effects of UV stellar spectral uncertainty on the chemistry of terrestrial atmospheres}

\correspondingauthor{Teal}
\email{teal@astro.umd.edu}

\author[0000-0002-1912-3057]{D.J. Teal}
\affiliation{Department of Astronomy, University of Maryland, College Park, MD 20742, USA}
\affiliation{NASA Goddard Space Flight Center}

\author[0000-0002-1337-9051]{Eliza M.-R. Kempton}
\affiliation{Department of Astronomy, University of Maryland, College Park, MD 20742, USA}

\author[0000-0003-2052-3442]{Sandra Bastelberger}
\affiliation{Department of Astronomy, University of Maryland, College Park, MD 20742, USA}
\affiliation{NASA Goddard Space Flight Center}
\affiliation{Center for Research and Exploration in Space Science and Technology, NASA/GSFC, Greenbelt, MD 20771}
\affiliation{Sellers Exoplanet Environments Collaboration, NASA Goddard Space Flight Center, Greenbelt, MD 20771, USA}

\author[0000-0002-1176-3391]{Allison Youngblood}
\affiliation{NASA Goddard Space Flight Center}
\affiliation{Sellers Exoplanet Environments Collaboration, NASA Goddard Space Flight Center, Greenbelt, MD 20771, USA}

\author[0000-0001-6285-267X]{Giada Arney}
\affiliation{NASA Goddard Space Flight Center}
\affiliation{Sellers Exoplanet Environments Collaboration, NASA Goddard Space Flight Center, Greenbelt, MD 20771, USA}



\begin{abstract}

The upcoming deployment of JWST will dramatically advance our ability to characterize exoplanet atmospheres, both in terms of precision and sensitivity to smaller and cooler planets. Disequilibrium chemical processes dominate these cooler atmospheres, requiring accurate photochemical modeling of such environments. The host star's UV spectrum is a critical input to these models, but most exoplanet hosts lack UV observations. For cases in which the host UV spectrum is unavailable, a reconstructed or proxy spectrum will need to be used in its place. In this study, we use the MUSCLES catalog and UV line scaling relations to understand how well reconstructed host star spectra reproduce photochemically modeled atmospheres using real UV observations. We focus on two cases; a modern Earth-like atmosphere and an Archean Earth-like atmosphere that forms copious hydrocarbon hazes. We find that modern Earth-like environments are well-reproduced with UV reconstructions, whereas hazy (Archean Earth) atmospheres suffer from changes at the observable level. Specifically, both the stellar UV emission lines and the UV continuum significantly influence the chemical state and haze production in our modeled Archean atmospheres, resulting in observable differences in their transmission spectra. Our modeling results indicate that UV observations of individual exoplanet host stars are needed to accurately characterize and predict the transmission spectra of hazy terrestrial atmospheres. In the absence of UV data, reconstructed spectra that account for both UV emission lines and continuum are the next best option, albeit at the cost of modeling accuracy. 

\end{abstract}

\keywords{editorials, notices --- 
miscellaneous --- catalogs --- surveys}


\section{Introduction} \label{sec:intro}

\subsection{Photochemical Modeling of Exoplanet Atmospheres}

The impending launch of the \textit{James Webb Space Telescope} (\textit{JWST}) will usher in an era of precise characterization of exoplanet atmospheres, including observations of smaller and cooler planets than ever before, and pushing ultimately toward the study of habitable worlds. \revision{With more precise observations on the horizon comes a requirement for improved models to meet the need of accurately interpreting future data sets.  One area that upcoming modeling efforts must address is that of disequilibrium chemistry.}{As more advanced observatories come online, the need for models able to accurately predict observations becomes increasingly crucial. Disequilibrium chemistry in particular has become important as modelling efforts strive for accuracy.}

Processes that drive an atmosphere away from chemical equilibrium include photochemistry and atmospheric mixing (both vertical and horizontal).  \revision{}{Such processes should impact planets across a range of parameter space.  Less irradiated planets are vulnerable to disequilibrium chemistry because chemical reactions that restore an atmosphere to thermochemical equilibrium are strongly temperature dependent and tend to proceed more slowly at cooler temperatures.  As such, the smaller and cooler exoplanets that will be uniquely observable with \textit{JWST} are expected to be significantly impacted by disequilibrium chemical effects, which therefore must be taken into account when modeling their atmospheres.  Conversely, hotter planets experience higher UV irradiation, commensurate with their higher instellation, and vertical mixing can be enhanced at higher temperatures \citep[e.g.][]{parmentier13}.  For hot Jupiters, the improved wavelength coverage and precision of \textit{JWST} observations will highlight departures from chemical equilibrium that may have been less apparent with prior data sets.}
\revision{Such processes should be more dominant for colder planets because chemical reactions that restore an atmosphere to thermochemical equilibrium are strongly temperature dependent and tend to proceed more slowly at cooler temperatures.  As such, the smaller and cooler exoplanets that will be uniquely observable with \textit{JWST} are expected to be impacted more significantly from disequilibrium chemical effects, which therefore must be taken into account when modeling their atmospheres.  Additionally, the improved wavelength coverage and precision of \textit{JWST} observations will highlight departures from chemical equilibrium that may have been less apparent with prior data sets.}{}

Chemical kinetics codes have been developed to predict the disequilibrium chemical composition of exoplanetary atmospheres by simultaneously tracking hundreds of chemical reaction rates and vertical mixing \citep[e.g.][]{line10, moses11,  kopparapu12, hu12, mil12, venot12, rimmer16, tsai17, kawashima18}.  \revision{Such}{The aforementioned} codes \revision{generally}{} use a numerical differential equation solver to enforce mass continuity throughout a one-dimensional (1-D) atmosphere, given production and loss rates within each vertically stratified layer and flux terms acting at layer boundaries.  Production and loss terms arise via the chemical reactions, and fluxes arise by processes such as eddy diffusion and molecular diffusion.  After setting physically appropriate boundary conditions at the top and bottom of the atmosphere, a steady-state solution is found by time stepping the solver forward until the chemical composition of each atmospheric layer remains stable at a predetermined threshold.  

Such chemical kinetics calculations have been applied to the study of exoplanet atmospheres to investigate the photochemical effects on atmospheric composition and aerosol production on a wide variety of exoplanet types, including hot Jupiters \citep[e.g.][]{zahnle09, line10, moses11, moses13, venot12}, hydrogen-rich Neptunes and sub-Neptunes \citep{line11, mil12, moses13b, venot16}, and terrestrial exoplanets orbiting a range of host stars \citep[e.g.][]{Segura2010, rug18, Wunderlich2019}. When taken together, these studies have confirmed the suspicion that cooler planets should \revision{}{generally} be more strongly impacted by disequilibrium chemistry and that such effects should be more readily observable with \textit{JWST}-quality spectra.

An important caveat is that photochemical hazes further complicate modeling predictions and observables. These optically thick particles, hydrocarbon and sulfurous hazes, are expected to become abundant below $\sim 850$ K in planetary atmospheres with conditions conducive to forming haze precursor molecules \citep{fortney13, morley15, gao20b}. Such hazes bring about significant departures from equilibrium chemistry solutions and are expected to have strong observable signatures \citep[e.g.][]{morley15, Arney2017, kawashima18, kawashima19}.

\subsection{Considering the Host Star's UV Spectrum}

The UV spectrum of a planet's host star is a critical input to chemical kinetics models.  It is the UV flux that establishes the rates of photolysis reactions and thereby governs a primary process that drives the atmosphere out of equilibrium at its upper boundary.  Unfortunately, the UV spectra of many exoplanet host stars have not been measured, which makes it difficult to accurately model the photochemistry occurring in their planets' atmospheres.  Currently, the \textit{Hubble Space Telescope} (\textit{HST}) is the only astronomical observatory capable of obtaining high resolution UV spectra of host stars between 1,000 and 3,000 \AA.  With \textit{HST} nearing the end of its lifetime and no comparable UV missions on the near-term horizon, there is a pressing need to identify which UV observations of exoplanet host stars must be obtained now to ensure future success in modeling and interpreting exoplanet spectra obtained with upcoming facilities like \textit{JWST}.  

With that in mind, considerable \textit{HST} observing time in recent years has been applied to UV monitoring of stars that are of particular interest to exoplanet studies. Notably, the MUSCLES\footnote{\url{https://archive.stsci.edu/prepds/muscles/}} Treasury Survey (\textit{HST} GO 13650, PI K.\ France) obtained UV observations with \textit{HST} of 12 M and K stars known to host exoplanets and used these to create high-resolution flux-calibrated panchromatic spectra \citep{France2016,Youngblood2016,Loyd2016}.  Later-type main sequence exoplanet hosts were targeted because UV observations of such stars were generally lacking, despite the fact that M stars offer the most favorable conditions for transit spectroscopy. Furthermore, the planets orbiting M stars are expected to be more highly impacted by the UV environment of their hosts \revision{because these}{which evolve over the course of a host star's lifetime \citep{Shkolnik2014,Luger2015}. Since later type} stars give off more UV radiation relative to their bolometric luminosities compared to earlier-type stars\revision{}{, this effect is particularly relevant to the most favorable targets for characterization of potentially habitable exoplanets}. Following on the MUSCLES survey, the Mega-MUSCLES survey \citep[\textit{HST} GO 15071, PI C.\ Froning;][]{froning19} expanded the sample of UV-characterized host stars to additional and even later-type stars, and various other UV studies of  exoplanet hosts and the M-dwarf population are being pursued as well \citep[e.g.][]{berta15, berta17, bourrier18, waalkes19, froning19b, france20, diamondlowe21, Loyd2021, Pineda2021}.

In the absence of observed UV data, various scaling relations have been defined to approximate a host star's UV spectrum based on optical proxies related to the Ca~II H \& K lines.  \citet{Youngblood2017} determined a scaling relation between the equivalent width of the Ca~II~K line and various UV emission lines using the UV spectra from the MUSCLES survey.  \citet{Melbourne2020} extended this work to also consider the full set of available \textit{HST} UV spectra of M dwarfs (a factor of $\sim7$ increase in sample size) and found that the $R'_{HK}$ index --- the Ca II H \& K line core intensity index defined in \citet{Rutten1984} --- was the best predictor of UV emission line strength of the observable proxies that they considered.  The advantage of using optical proxies is that, in principle, a star's UV spectrum can be approximated in the absence of observed UV data using information readily accessible to ground-based observatories.  The Ca II H \& K lines, at 3969 \AA\ and 3934 \AA, respectively, are historically well-observed, and the $R'_{HK}$ index has been cataloged for many stars or can otherwise be calculated from existing optical spectra. 

For models requiring UV spectra as inputs, such as photochemical models, determining if these reconstructions are sufficient in the absence of observations allows informed decision-making when choosing stars to observe before \textit{HST} is unavailable. This work aims to close the loop on that question with respect to photochemical modeling in particular. We do this by directly comparing the outputs of photochemical models run using observed stellar spectra vs.\ those run using the \citet{Melbourne2020} reconstructions of the same UV spectra. We then examine the degree to which the transmission spectra of each of the modeled exoplanets are altered by the use of the UV reconstructions, and we comment on implications for interpreting observations from \textit{JWST}. In Section~\ref{sec:methods}, we describe our photochemistry-climate model, the UV observations that we use as inputs to this model, and how we reconstruct the UV spectra of our input stars. In Section \ref{sec:photochemistry results} we describe the results of our photochemical models, with discussion of their resulting transmission spectra in Section \ref{sec:transmission spectra results}. Section \ref{sec:continuum treatment results} explores the impact of various reconstructions of the UV continuum (rather than the UV emission lines), focusing on the host star GJ 176.  Finally, Section \ref{sec:conclusion} summarizes this work and offers discussion of the implications of our results, as well as motivation for future study.

\section{Methods} \label{sec:methods}

\subsection{Photochemistry and climate model} \label{sec:atmos}

We use the \Atmos coupled 1-D photochemistry and climate model to simulate the physical properties of all atmospheres in our study. This model is well-established in the literature, having been used to investigate the effects of stellar activity on Earth-like atmospheres \citep[e.g.][]{Segura2005,Segura2010}, hazy terrestrial (``Archean'') atmospheres \citep[e.g.][]{Arney2016,Arney2017,Fauchez2019}, and numerous other studies of Earth-like atmospheres under various conditions \citep[e.g.][]{Kasting1979,Kopparapu2017,Harman2018,Meadows2018,Afrin-Badhan2019}. 

\Atmos includes the option to iterate between a photochemical model and a 1-D climate (radiative-convective equilibrium) model until reaching a steady-state solution. \Atmos' photochemical model includes a variety of important physical processes\revision{, }{---}such as lightning \citep{Harman2018}, haze formation \citep{Arney2016}, sedimentation and rainout \citep{Arney2017}\revision{;}{---}in addition to standard gas-phase chemical and photolysis reactions. This \revision{fully-functional}{extensively developed and recently updated} chemical kinetics model works in conjunction with the included climate model, which determines the 1-D temperature-pressure profile in radiative-convective equilibrium, to explicitly model temperature-sensitive processes such as water saturation and humidity \citep{Kasting1986,Mischna2000,Kopparapu2013}. These coupled models allow for feedback between radiative-convective equilibrium and chemical steady-state self-consistently.

\Atmos provides two well-tested atmospheric templates that we use in this study: an oxygen-rich, hazeless\footnote{Technically, sulfur aerosols are included in the modern Earth template, but their abundances are trace, and they do not significantly alter the atmospheric state.} modern Earth-like template and a hazy Archean Earth template suitable for low oxygen conditions. These serve as the initial conditions for each of our simulations. Table \ref{tab:boundary conditions} shows the boundary conditions for each chemistry model, which include species specific to either atmospheric state. For our modern Earth and Archean models, we use updated and extended versions of the reaction networks described in \citet{Afrin-Badhan2019}, \citet{Lincowski_2018} and \citet{Arney2017}, with the most up-to-date version of these reaction networks appearing in the \Atmos GitHub repository\footnote{\href{https://github.com/VirtualPlanetaryLaboratory/atmos}{https://github.com/VirtualPlanetaryLaboratory/atmos}}.

\begin{deluxetable*}{l@{\hskip 0in}l@{\hskip 0in}l||l@{\hskip 0in}l@{\hskip 0in}l}
\label{tab:boundary conditions}
\tablehead{\colhead{Species} & \colhead{Type} & \colhead{Value} & \colhead{Species} & \colhead{Type} & \colhead{Value}}

\startdata
\textbf{Both models} \phs& \phs& \phs&
    \textbf{Modern Earth} \phs& \phs& \\ \hline
 \ce{O} \phs&  Deposition velocity \phs&  1.0 \phs&
    \gcell \ce{O2} \phs& \gcell Fixed Mixing Ratio \phs& \gcell $2.1 \times 10^{-1}$\\
\ce{H} \phs& Deposition velocity \phs& 1.0 \phs& 
    \gcell \ce{H2} \phs& \gcell Fixed Mixing Ratio \phs& \gcell $5.3\times10^{-7}$\\
\gcell \ce{OH} \phs& \gcell Deposition velocity \phs& \gcell 1.0 \phs&
    \gcell \ce{CO} \phs& \gcell Flux \phs& \gcell $3.7\times10^{11}$\\
 \ce{HO2} \phs&  Deposition velocity \phs&  1.0 \phs&
    \gcell \ce{CH4} \phs& \gcell Flux \phs& \gcell $1.0\times10^{11}$\\
 \ce{H2O2} \phs&  Deposition velocity \phs&  $2.0 \times 10^{-2}$
    \phs& \ce{N2O} \phs& Flux \phs& $1.53\times10^{9}$\\
\ce{HCO} \phs& Deposition velocity \phs& 1.0 \phs&
    \gcell \ce{H2S} \phs& \gcell Flux \phs& \gcell $1.0\times10^8$\\
\gcell \ce{H2CO} \phs& \gcell Deposition velocity \phs& \gcell $2.0 \times 10^{-1}$ \phs&
    \ce{HO2NO2} \phs& Deposition Velocity \phs& $0.2$\\ \cline{4-6}
\gcell \ce{NO} \phs& \gcell Deposition velocity \phs& \gcell $3.0 \times 10^{-4}$ \phs& 
    \textbf{Archean Earth} \phs& & \\ \cline{4-6}
\gcell \ce{NO2} \phs& \gcell Deposition velocity \phs& \gcell $3.0 \times 10^{-3}$ \phs&
    \gcell \ce{H2} \phs& \gcell Deposition Velocity \phs& \gcell $2.4\times10^{-4} $\\
\gcell \ce{HNO} \phs& \gcell Deposition velocity \phs& \gcell 1.0 \phs&
    \gcell \phs& \gcell Flux \phs& \gcell $1.0\times10^{10}$ \\
\gcell \ce{H2S}\tablenotemark{a} \phs& \gcell Deposition velocity \phs& \gcell $2.0 \times 10^{-2}$ \phs&
    \gcell \ce{O2} \phs& \gcell Deposition Velocity \phs& \gcell $1.0\times10^{-4}$ \\
\gcell \ce{SO2} \phs& \gcell Deposition velocity \phs& \gcell 1.0 \phs&
    \gcell \ce{CO} \phs& \gcell Deposition Velocity \phs& \gcell $1.2\times10^{-4}$ \\
 \gcell \phs& \gcell Flux \phs& \gcell $1.0 \times 10^9$ \phs&
    \gcell \ce{H2S} \phs& \gcell Flux \phs& \gcell $3.5\times10^{8}$\\
\ce{H2SO4} \phs& Deposition velocity \phs& 1.0 \phs&
    \gcell \ce{CH4} \phs& \gcell Fixed Mixing Ratio \phs& \gcell $3.5\times10^{-3}$\\
\ce{HSO} \phs& Deposition velocity \phs& 1.0 \phs& 
    \gcell \ce{C4H2} (Aerosol) \phs& \gcell Deposition Velocity \phs& \gcell $1.0\times10^{-2}$\\
\ce{SO4} (Aerosol) \phs& Deposition velocity \phs& $1.0 \times 10^{-2}$ \phs&
    \gcell \ce{C5H4} (Aerosol) \phs& \gcell Deposition Velocity \phs& \gcell $1.0\times10^{-2}$\\
\ce{S8} (Aerosol) \phs& Deposition velocity \phs& $1.0 \times 10^{-2}$ \phs&
    \gcell \ce{CO2}  \phs& \gcell Fixed Mixing Ratio \phs& \gcell $2.0\times10^{-2}$\\
    \gcell \ce{O3} \phs& \gcell Deposition velocity \phs& \gcell $7.0 \times 10^{-2}$ \\
\ce{CH3} \phs& Deposition velocity \phs& 1.0 \phs&
    &&\\
\ce{HNO3} \phs& Deposition velocity \phs& $2.0 \times 10^{-1}$ &&&\\
\enddata
\caption{Static boundary conditions at the surface of our model. Deposition velocity has units of cm/s. Flux is a constant surface flux of a species measured in molecules/cm$^2$/s. We note that this does not include top-of-atmosphere fluxes, such as downward fluxes of \ce{CO} and \ce{O}, which are parameterized in the model based on abundances at the top of the atmosphere \citep{Arney2016,Afrin-Badhan2019}.  Species highlighted in gray are those that are included as opacity sources in our \texttt{Exo-Transmit} calculations (Section~\ref{sec:ExoTransmit}).}
\tablenotetext{a}{\ce{H2S} deposition is an additional boundary condition alongside the fluxes  and is the same across both model templates. This flux, along with fluxes of \ce{SO2} and \ce{H2}, are distributed within the troposphere and meant to account for volcanic outgassing in both models.}
\end{deluxetable*}

The latest public version of the \Atmos model includes several significant updates from previously published versions. The changes relevant to this study are summarized below:
\begin{itemize}
\item For the climate model, the $k$-coefficients for \ce{H2O} and \ce{CO2} were updated using the HITRAN2016 database \citep{GORDON20173}. For \ce{H2O}, we assume 25 cm$^{-1}$ line cut-offs using Lorentz profiles with the plinth removed. For \ce{CO2}, we use 500 cm$^{-1}$ line cut-offs using the Perrin and Hartman sub-Lorentzian line profiles (\cite{Perrin_1989}; standard values for coarse spectral resolution). The coefficients were generated using \texttt{HELIOS-k}\xspace \citep{Grimm_2015}. 
\item The photochemical model uses a 750 bin wavelength grid --- the same one from \citet{Lincowski_2018} spanning $1176.5 - 10000$ \AA \space with a resolution of 100 cm$^{-1}$. This grid resolves the UV-wavelength range critical to this study significantly better than the previous 118-bin grid. \revision{}{In particular, important UV lines such as \lya are no longer spread over a broad wavelength range. Instead, these lines are better-resolved, resulting in more accurate calculations of photolysis rates for all molecules. When spread over a wider wavelength range as in the previous wavelength grid, molecules that are otherwise not sensitive to a given strong line will have an overestimated photolysis rate. Similarly, species very sensitive to these lines will have an underestimated photolysis rate.}
\item Comprehensive updates have been made to the photolysis cross sections and quantum yield data for the photochemical model \citep[e.g.\ \ce{H2O} cross sections from][]{Ranjan_2020}. The updated cross section and quantum yield data were sourced from \cite{HEBRARD2006211}, \cite{Lincowski_2018}, the JPL Publication 19-5 recommendations \citep{burkholder2019} and the MPI-Mainz UV/VIS Spectral Atlas  \citep{MPI_spectral_atlas_2013}; and references therein.

\item The treatment of hydrocarbon aerosols has been updated such that different production channels now all contribute to a single particle population. Previously, each production pathway formed non-interacting, distinct particle populations with only one pathway providing opacity in the climate model while the contribution of other pathways was neglected.

\item We include new options for hydrocarbon aerosol optical constants, such as new UV-visible refractive index data for early Earth aerosols \citep{Gavilan2017}, which are used for the Archean models in this study, and different monomer sizes (ranging from 10 to 70 nm) for fractal particles. 
 
\end{itemize}

In addition to the list above, we have also implemented an updated convergence scheme for coupled photochemistry-climate models involving significant haze formation (i.e.\ our Archean Earth models).  In contrast to an integrated model that solves both photochemistry and radiative transfer simultaneously,  \Atmos relies on external coupling of historically separate climate and photochemistry models --- the two models are run sequentially in an iterative fashion. When the \Atmos model is run in this manner, the external coupling between the two models may impede a self-consistent atmospheric solution in some cases. Feedbacks between molecular and/or aerosol abundances and the thermal state of the atmosphere can cause the coupled model to oscillate between two non self-consistent solutions with drastically different temperature-pressure and chemical profiles. For example, hazy states can lead to significant atmospheric heating, which in turn will destroy hazes on a subsequent model run --- thus impeding overall model convergence.  To avoid this problem, we use a ``short-stepping'' method, in which we do not allow the climate model to fully adjust the temperature-pressure profile to the radiative forcing exerted by the spectrally active species, and instead we interrupt the climate code after a limited number of iterative steps before re-calculating changes to the chemistry with the photochemistry model. Over many iterations of the coupled code, this allows for more reliable convergence to a self-consistent atmospheric steady-state solution in radiative-convective equilibrium.  Only in the final step of a coupled model run, once the thermal structure and chemical composition of the atmosphere appear to have settled into a stable state, do we finally allow the climate model to run to a converged solution.  

\subsubsection{Modelling aerosols} \label{sec:haze model}

Hydrocarbon aerosol particles are thought to have intermittently existed in Earth's atmosphere during the Archean period \revision{}{\citep[e.g.,][]{Zerkle2020}}. Such hazes are modeled in our Archean Earth template, replicating a complex mixture of massive molecules with distinct optical properties compared to gas-phase molecular species. These particles are thought to exist in a variety of atmospheric types and planetary conditions \citep{Horst2018,Fleury2019b,Zerkle2020}. 

The formation of hydrocarbon haze is initiated by the photolysis of \ce{CH4} and then proceeds via complex and poorly understood chemical polymerization pathways.  As a result, it is not feasible to model the entire chemical reaction network leading to haze production, and we instead follow a common modeling practice of \revision{directly}{} converting certain high-order gas-phase hydrocarbon molecules directly into insoluble haze \citep{Pavlov2001,Lavvas_2008_1,Krasnopolsky_2009,morley15,Arney2016}.  For the purposes of our model we assume two high-order hydrocarbon species will ultimately condense into haze particles with a 100\% conversion efficiency. These ``haze precursors'' are \ce{C4H2} and \ce{C5H4}, formed via the reactions:

\begin{align} \label{eq:haze rxns}
\begin{split}
    &\ce{C2H + C2H2 -> C4H2 + H} \\
    &\ce{C2H + CH2CCH2 -> C5H4 + H}
\end{split}
\end{align}

Large particles scatter very efficiently, introducing significant opacity to an atmosphere, which \revision{obfuscates}{obscures} the spectral features of other molecules and fundamentally alters the thermal balance throughout an atmosphere \citep{Arney2016,Arney2017,Arney2018,Lavvas2021}.
The refractive index of \revision{experimentally produced}{experimentally-produced} aerosol condensate is influenced by the chemical composition of the gas mixture in which it was produced, and may also be dependent on the energy source used to generate the particles \citep{Hasenkopf2010,Mahjoub2012,Gavilan2017,2018Gavilan,He2018,Ugelow_2018}. \citet{Gavilan2017} found enhanced UV absorption in oxidized aerosol material produced in early Earth-like  \ce{N2}/\ce{CO2}/\ce{CH4} mixtures compared to more reducing mixtures. The real and imaginary part of the refractive index may exhibit strong wavelength dependence, and data covering the whole UV/Vis/IR range is rarely available, with few exceptions \citep{Sagan1984}. 

To contend with these complications, the optical properties of haze particles require specific treatments in our modeling in order to adequately capture scattering, absorption, and emission from particles that consist of agglomerations of hydrocarbons with non-uniform chemical makeup. Several approaches have been used in previous work, including enhanced Rayleigh scattering and Mie scattering approximations. In this work, we follow the approach laid out in \cite{Arney2016}, which is to treat the hazes as fractal aggregates using the mean-field approximation \citep{Botet1997, Rannou1997}. Further, we apply the refractive indices of \citet{Gavilan2017} for early Earth like atmospheres (\ce{N2}:\ce{CO2}:\ce{CH4}= 90:8:2) in the UV-visible range over which they were reported, and \citet{Sagan1984} for the IR.  

\begin{figure}[ht]
    \centering
    \includegraphics[width=3.35in]{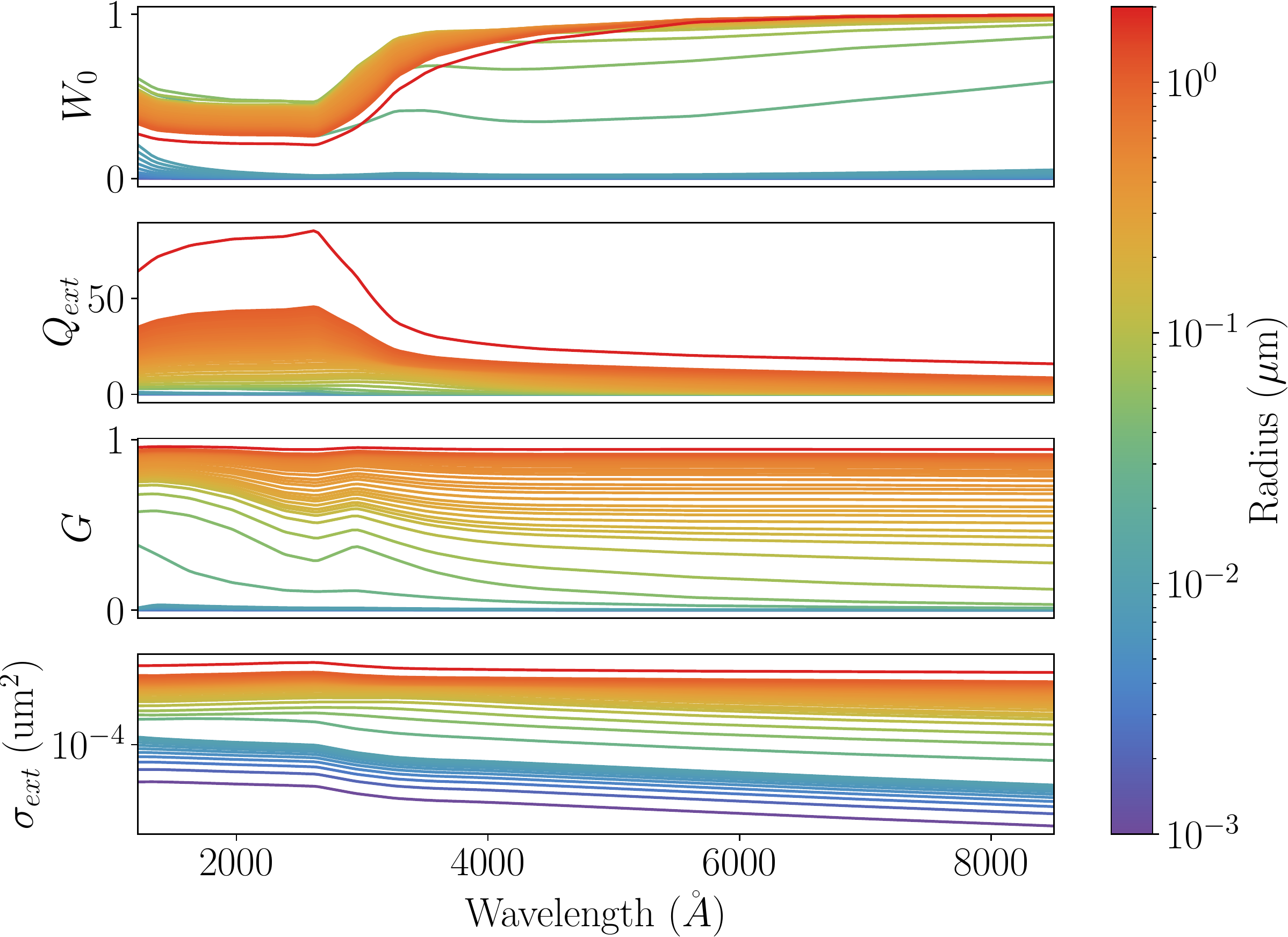}
    \caption{Optical properties for fractal haze (radius $>$ 50 nm) and spherical monomer (radius $<$ 50 nm) particles across radii modeled the photochemistry and climate models. $W_0$ is the single scattering albedo, $Q_{ext}$ is the extinction efficiency, $G$ is the asymmetry parameter, and $\sigma_{ext}$ is the effective haze extinction cross section for transmission spectroscopy. These are the same optical properties employed in both the \Atmos and \texttt{Exo-Transmit} models. These optical properties are calculated using the fractal haze model described in \citep{Rannou1997} and \citep{Botet1997}, with haze optical properties from \cite{Gavilan2017} and \cite{Sagan1984}.}
    \label{fig:haze cross sections}
\end{figure}

The mean-field approximation considers the effects of non-spherical haze particles on radiative transfer through the atmosphere. Given refractive indices for a non-spherical particle as a function of wavelength, we can then calculate the extinction coefficient $Q_{ext}$, single scattering albedo $W_0$, and asymmetry parameter $G$ (Figure~\ref{fig:haze cross sections}). These scattering parameters are employed in the two-stream radiative transfer calculations in both the \texttt{Atmos} photochemistry routines (for UV wavelengths) and climate routines (for visible / IR wavelengths) to account for multiple scattering off of aerosol particles.  These same scattering parameters are also used in our transmission spectroscopy radiative transfer to calculate an effective extinction cross section $\sigma_{ext}$, further described in Section \ref{sec:ExoTransmit} (see Equation \ref{eq:haze cross section}).

As shown in Figure~\ref{fig:haze cross sections}, we use a grid of haze optical properties spanning radii of 1 nm to 2 $\mu$m over wavelengths between 1216 \AA\  and 9000 \AA. In our model, the haze particle radius is determined based the coagulation time scale and removal times scales through diffusion and sedimentation at a given pressure level \citep{Arney2016}. Particles are first treated like spherical Mie scatters as they grow from nucleation size to a size of 50 nm, after which they are considered fractal aggregates comprised of spherical monomers. \revision{}{This threshold is chosen to represent previous work done to understand the haze properties within the Archean Earth and Titan's atmosphere, though we also find that our results are robust to other choices in initial particle sizes \citep{Tomasko_2008, Larson_2015}.} 
As hazes form and interact with the local radiation field, they can significantly alter the thermal balance of an atmosphere. \revision{Arney et al. (2017)}{Previous studies have shown} that increasing haze abundance significantly warms high altitudes where they form, while simultaneously cooling the planet's surface \revision{}{\citep[e.g.,][]{Pavlov2001,Arney2017,Lavvas2021}}. Furthermore, haze properties, particularly for larger particles, are sensitive to small changes in temperature \citep{Horst2018}, resulting in haze abundance, particle radii, and formation rates varying non-linearly over different temperatures and levels of irradiation.  The use of the \texttt{Atmos} coupled climate and chemistry models allows us to model and account for these sensitive feedbacks between haze properties and the thermal structure of the atmosphere.

\subsection{UV input spectra for photochemical modeling} \label{sec:MUSCLES}

\begin{figure*}[t]
    \centering
    \includegraphics[width=6.5 in]{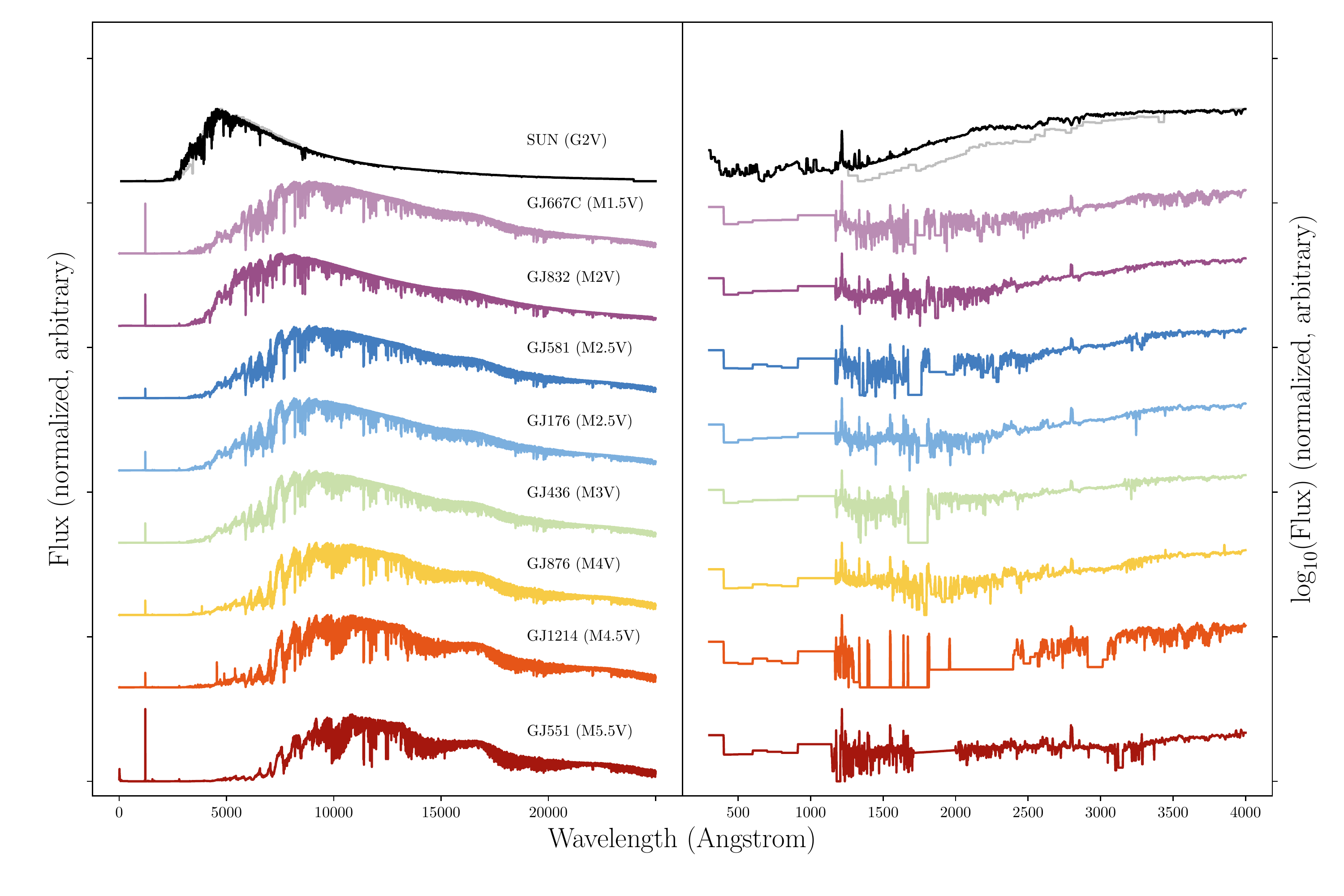}
    \caption{The spectra used in this work. The vertical axis represents the normalized stellar flux, with an arbitrary offset applied.  When running \Atmos, each spectrum is re-scaled such that the total insolation is Earth-like for the modern-Earth and Archean simulations, with the exception of the solar Archean model. 
     \citep{Claire_solar}\revision{For the solar Archean model, we use 0.813 times the modern Earth insolation to account for a fainter young sun 2.7 billion years ago}{For our Archean models using the solar spectrum as input, a model following \citet{Claire_solar} is used to account for predicted differences in the solar spectrum 2.7 billion years ago. This treatment is not applied to our M-dwarf models.}.  The horizontal axis is wavelength, with the left panel being the full panchromatic MUSCLES spectrum (in linear flux units) and the right being a zoom-in on the UV wavelength range over which we reconstruct the spectrum (in log flux units).}
    \label{fig:all_specs}
\end{figure*}

In this study, we focus on the validity and accuracy of reconstructed UV spectra as input to photochemical models. To that end, we select and reconstruct stellar UV spectra using the following methodology.

\begin{figure*}[t]
    \centering
    \includegraphics[width=7in]{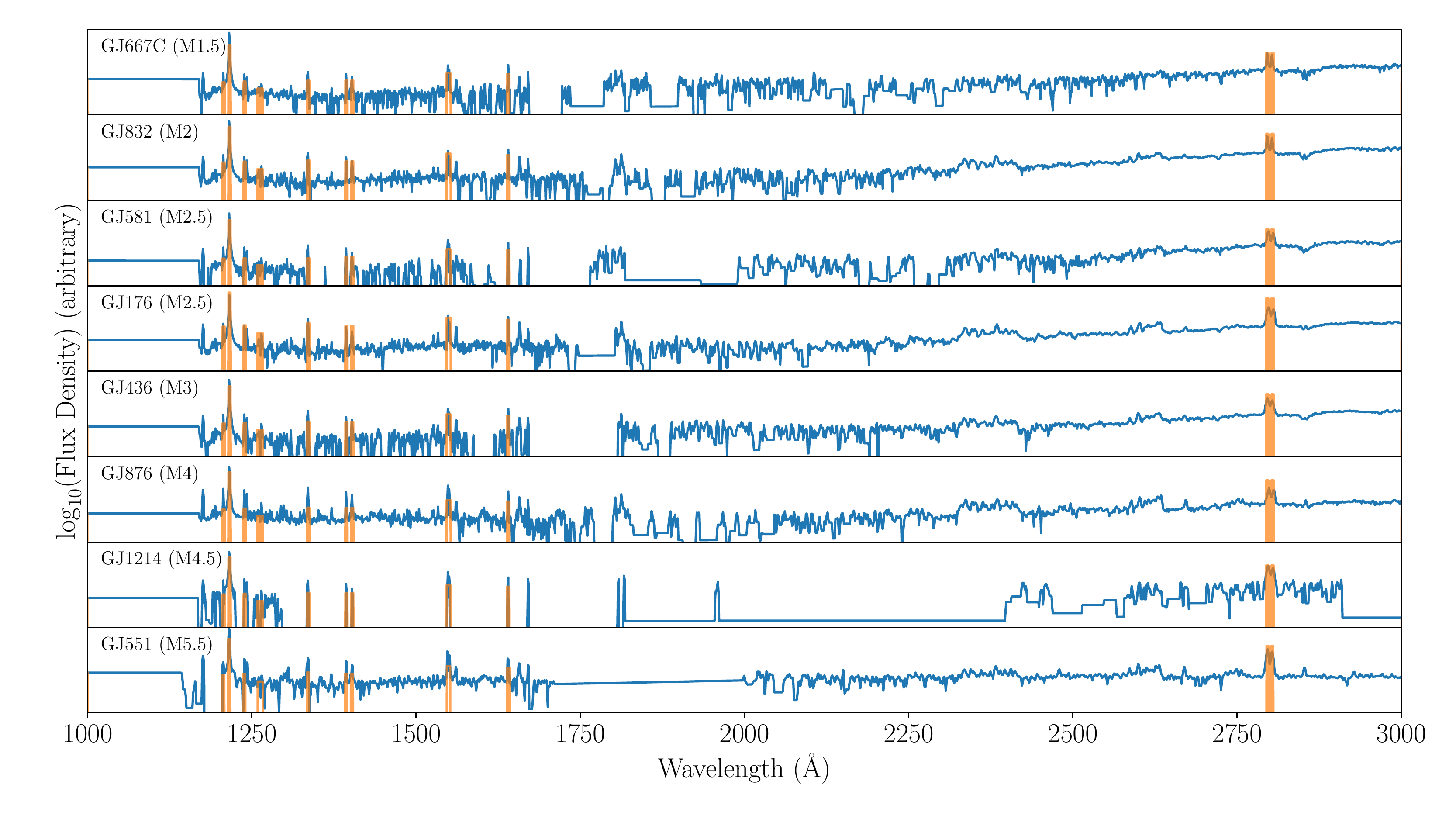}
    \caption{MUSCLES spectra (blue) overplotted with our line reconstructions (orange) using the \citet{Melbourne2020} scaling relations. The reconstructed line profiles are 2-\AA\ \revision{boxcars}{top-hat profiles}, as described in the text.  The reconstructed spectra shown here have our zero-continuum treatment applied.}
    \label{fig:line reconstructions}
\end{figure*}

\begin{figure*}[t]
    \centering
    \includegraphics[width=6in]{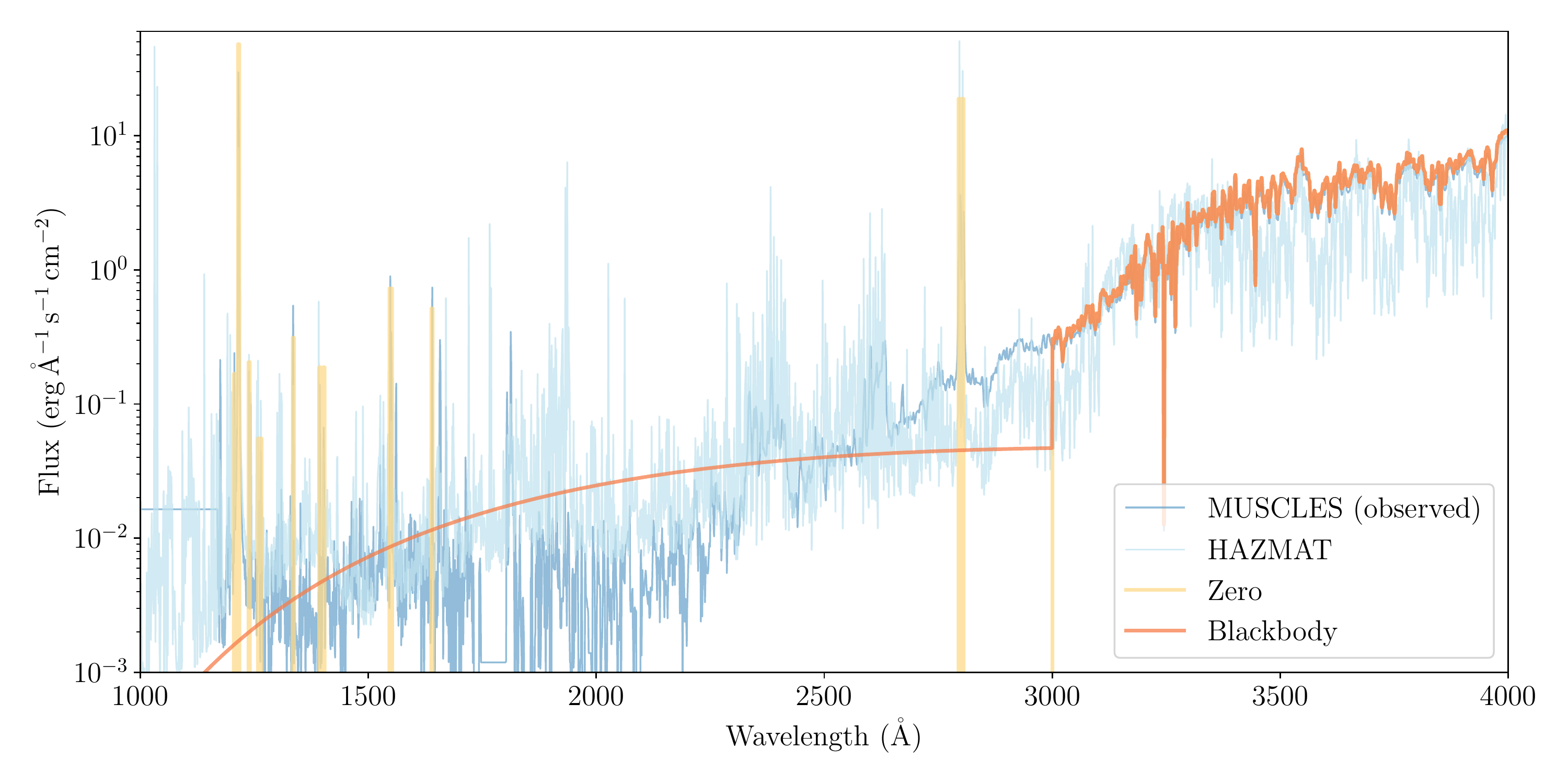}
    \caption{Input GJ 176 UV spectra for several of the tested continuum treatments. All cases plotted, save for the baseline observed MUSCLES spectrum case, have the same set of reconstructed emission lines as described in Section \ref{sec:MUSCLES}.  We additionally run a final set of two Archean Earth models using the continua of GJ 581 (M2.5) and GJ 436 (M3) in place of GJ 176 (M2.5).}
    \label{fig:gj176 all continua}
\end{figure*}

\begin{deluxetable*}{l@{\hskip 0in}l@{\hskip 0in}l l@{\hskip 0in}l@{\hskip 0in}l}
\label{tab:stellar information}
\tablehead{\colhead{Star name} & \colhead{Type} & \colhead{Radius ($R_{\cdot}$)} & \colhead{$T_{eff}$ (K)} & $\log_{10}(R'_{HK})$} 

\startdata
Sun\tablenotemark{a} & G2 & 1. & 5800\\
GJ 667c & M1.5 & 0.46 & 3450 & -5.47\\
GJ 832 & M2 & 0.56 & 3590 & -5.22\\ 
GJ 581 & M2.5 & 0.3 & 3500 & -5.75\\
GJ 176 & M2.5 & 0.45 & 3680 & -4.89\\
GJ 436 & M3 & 0.45 & 3420 & -5.45\\
GJ 876 & M4 & 0.38 & 3130 & -5.48\\
GJ 1214 & M4.5 & 0.21 & 2820 & -5.47\\
GJ 551 (Proxima Centauri) \phs& M5.5 & 0.14 & 3100 & -5.23\\ 
\enddata

\caption{Table of MUSCLES stars used in this work. Each of the $R'_{HK}$ values are taken from \citet{Melbourne2020} and the citations therein. $T_{eff}$ is the star's effective temperature, and $R'_{HK}$ is the Ca II H \& K line core intensity index.}
\tablenotetext{a}{The solar spectrum is not from the MUSCLES catalog, and instead is the default solar spectrum shipped with the \Atmos photochemistry model. The model also scales this spectrum based on the age of the Solar System being used, but we do not scale any other spectra with age.}
\end{deluxetable*}

First, we create a set of baseline photochemical models using panchromatic spectra from the MUSCLES Treasury Survey \citep{France2016,Loyd2016,Youngblood2016}. We use the adaptive, constant-resolution data products to avoid overestimation of flux when handling negative flux bins and re-binning to the \texttt{Atmos} wavelength grid\footnote{See the \href{https://archive.stsci.edu/missions/hlsp/muscles/hlsp_muscles_multi_multi_all_broadband_v22_readme.pdf}{MUSCLES documentation} at \href{https://archive.stsci.edu/prepds/muscles/}{https://archive.stsci.edu/prepds/muscles/}.}. This sample contains 8 M-stars with spectral types ranging from M1.5 to M5.5. In addition, as our 9th host star, we run baseline models using the solar spectrum template included in the \texttt{Atmos} code. Figure \ref{fig:all_specs} depicts the normalized spectra for all of the MUSCLES M-stars, as well as the Sun.  In the following analysis, we treat these spectra as the ``ground truth'' for the stellar UV, though we discuss the nuances of this assumption in Section \ref{sec:correct continuum treatment}. Our baseline model grid is made up of 18 individual \texttt{Atmos} runs: models at both modern Earth and Archean Earth initial conditions are produced for each of the 9 host stars. Table~\ref{tab:stellar information} provides a list of stars used in this work as well as properties relevant to each star's UV spectrum reconstruction and transmission spectrum calculations. 
 
Next, we regenerate each of our photochemical models using reconstructed MUSCLES UV spectra obtained by applying the UV line scaling relations described in \citet{Melbourne2020}.  Specifically, these scaling relations estimate a given line luminosity using the following equation:
\begin{equation}
    \label{eq:scaling relation}
    \log_{10}(L_{UV}/L_{bol}) = \alpha \log_{10}(R'_{HK}) + \beta
\end{equation}
where $R'_{HK}$ is the Ca H \& K line core intensity index \citep{Rutten1984}, $\alpha$ and $\beta$ are fit parameters given in \citet{Melbourne2020}, and $L_{UV}$ and $L_{bol}$ are the UV line luminosity and the star's bolometric luminosity, respectively.  Using the values for $\alpha$, $\beta$, and $R'_{HK}$ from \citet{Melbourne2020}, we reconstruct each of 10 UV emission lines using Equation~\ref{eq:scaling relation}.  The reconstructed line profiles are taken to be top-hat\revision{s}{ profiles} with 2-\AA\ width (filling two adjacent bins in the 1 \AA-resolution input spectrum grid), centered on the line core, and with total wavelength-integrated luminosity equivalent to $L_{UV}$. Because the \citet{Melbourne2020} scaling relations are only for the strongest UV lines, and because most of the stellar UV flux emanates from these emission lines, we initially make the simplifying assumption of zero UV continuum flux outside of the emission line wavelength ranges. \revision{}{Furthermore, the \citet{Melbourne2020} relations are a linear regression of over 24 M-dwarf stars.  Due to intrinsic scatter in line intensities across the sample, uncertainties arise in these relations. In this work, we take the reported scaling relation parameters at face-value. We have performed limited tests to assure that our results do not differ significantly when accounting for 1-$\sigma$ scatter in these scaling parameters.}

Figure \ref{fig:line reconstructions} shows the full set of UV reconstructed spectra overlaid on the MUSCLES spectra.  In total, our full set of UV reconstructed photochemistry models consists of 16 individual \texttt{Atmos} runs --- one for each of the MUSCLES M-stars at both modern Earth and Archean Earth initial conditions.  These UV reconstructed spectra effectively simulate a situation in which no observed UV data are available for a given exoplanet host star.  This most basic reconstruction --- i.e.\ completely ignoring any possible continuum flux --- only has appreciable flux at the reconstructed lines. The remaining wavelengths are set to a constant value of $10^{-50}$ erg/cm$^2$/s/\AA, which is vanishingly small but nonzero to avoid numerical instabilities when running the \Atmos code.

Finally, we run a subset of models designed to quantify the impact of the UV continuum treatment on our results. In these cases we focus on the star GJ~176 (M2.5V), which is a representative early M-dwarf from the MUSCLES sample. In addition to the zero-continuum reconstructions described above, we examine three other approaches for reconstructing the UV continuum. In the three cases described below, and shown in Figure \ref{fig:gj176 all continua}, the continuum treatment is applied at all UV wavelengths (5 - 4000 \AA) other than those of the reconstructed UV emission lines, which are produced using the procedure already described above.  
\begin{enumerate}
    \item Blackbody continuum flux --- Because a zero continuum level is certainly an underestimate of the true UV emission, we employ a first approximation of a blackbody UV continuum at a temperature of 9000 K.  We select this temperature to compensate for increased UV flux not captured by a blackbody of an M dwarf's effective temperature. This approach na\"ively assumes the bulk of continuum flux originates from thermal radiation from plasma in the upper chromosphere \citep{Ayres1979,France2013,Peacock2019a}. We normalize our blackbody spectrum such that the total UV flux is equal to the total UV flux for GJ176 in the same wavelength range, minus flux contributions from the lines we reconstruct. \revision{}{Chromospheric temperatures can vary by several thousand degrees \citep{Mauas1997}, which will change the continuum flux from chromospheric emission appreciably. We choose a 9000 K continuum flux value to roughly follow the continuum flux exhibited in the MUSCLES data for GJ 176 as a test case based on observations.}
    
    \item Observed continuum flux --- In this approach, we retain the continuum flux recorded by the observed MUSCLES spectra and stitch this together with the reconstructed UV emission lines.  The goal here is to quantify how much of the photochemistry is being caused by the observed UV continuum vs.\ the strong (reconstructed) emission lines.
    
    \item Synthetic continuum flux --- Here we replace the continuum with a model UV spectrum. Specifically, we use the HAZMAT semi-empirical model spectra for this set of continuum reconstructions \citep{Peacock2019b}, which provide panchromatic spectra generated by the PHOENIX stellar atmospheric code and informed by GALEX and HST observations.
    
    \item Adjacent spectral type --- With this method, we take the observed continuum flux from a star of a neighboring stellar type and reconstruct the lines given by the scaling relations.
\end{enumerate}

Figure~\ref{fig:gj176 all continua} shows the reconstructed GJ 176 spectra using the first three continuum treatments from the list above. These comprise a set of 6 additional photochemical models --- one for each continuum treatment at both modern Earth and Archean Earth conditions.  We additionally apply the final continuum treatment --- employing the continuum of an adjacent spectral type --- to the Archean Earth model only, for two different adjacent host star spectra (GJ 581 and GJ 436).  

\subsection{\texttt{Exo-Transmit} transmission spectra \label{sec:ExoTransmit}}

The \texttt{Exo-Transmit} code \citep{Kempton2017} is used to generate transmission spectrum observables for each of our model atmospheres. The version of \texttt{Exo-Transmit} we use has been modified from the original code to accept the non-equilibrium, vertically-defined chemical abundance profiles output by \texttt{Atmos}, rather than the equilibrium chemistry models provided.  This modification consists of a major overhaul to the ordering in which chemistry, opacity, and optical depth data are read in and calculated within the code but otherwise leaves the transmission spectrum calculation unchanged.  

Mixing ratio profiles for species output by \Atmos and shaded gray in Table~\ref{tab:boundary conditions} are read into \texttt{Exo-Transmit}, in addition to \ce{C2H2}, \ce{C2H4}, \ce{C2H6}, \ce{OCS}, \ce{NH3}, and \ce{HCN}, which \revision{do not have specified}{have null} boundary conditions in our models (and therefore do not show up in Table \ref{tab:boundary conditions}), but \revision{can}{} form through reactions. The molecular opacities we use for all species are those included in the public \texttt{Exo-Transmit} GitHub repository\footnote{\url{https://github.com/elizakempton/exo_transmit}} and documented in \citet{Kempton2017}. The total opacity for each layer is determined by geometrically weighting the individual species' opacities by their respective mixing ratios in each vertical layer of the atmosphere.

Our hazy (Archean) model runs must also include contributions of hydrocarbon haze particles to the opacity of the atmosphere. To accomplish this, we include the hydrocarbon aerosols as an additional extinction species in \texttt{Exo-Transmit} using the following procedure. We first calculate haze extinction cross sections $\sigma_{ext}$ as a function of particle radius, $r_{par}$, according to \begin{equation} 
\sigma_{ext} = \pi r_{par}^2 Q_{ext} (1 - G^2)
\label{eq:haze cross section}
\end{equation}
where $Q_{ext}$ is the extinction efficiency, and $G$ is the asymmetry parameter.  (The final term in this equation is a correction based on the asymmetry parameter to account for the fraction of incoming starlight that is forward scattered and therefore remains in the beam.)  For each atmospheric layer, the total haze opacity (in units of m$^{-1}$) is obtained by selecting $\sigma_{ext}$ at the nearest neighbor to the mean particle radius in that layer and then multiplying by the haze number density output by \Atmos.  We use the same wavelength-dependent haze optical properties from \Atmos (Figure~\ref{fig:haze cross sections}) for self-consistency between our \Atmos and \texttt{Exo-Transmit} calculations.

\section{Photochemical modeling results}
\label{sec:photochemistry results}
\begin{figure*}[t]
    \centering
    \includegraphics[width=\textwidth]{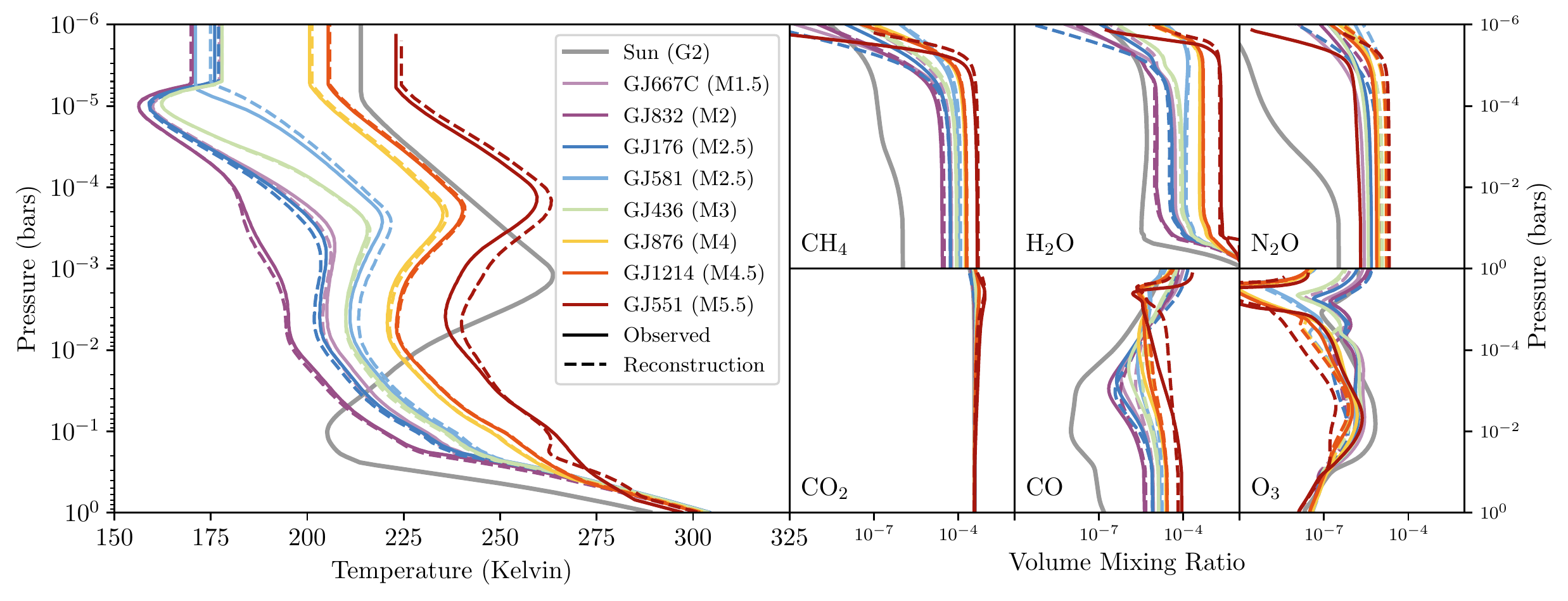}
    \caption{Modern Earth model results for our baseline case using the MUSCLES observations (solid curves), and the same models re-run with the (zero-continuum) \citet{Melbourne2020} UV spectral reconstructions (dashed lines). These models are effectively haze-free.  Vertical temperature-pressure profiles are plotted to the left, whereas mixing ratio profiles for various species (as indicated) are plotted in the right-hand panels. Thermal inversions for the M-dwarf models at $\sim 10^{-3}$ bar are caused by \ce{H2O}, unlike Earth's inversion caused by \ce{O3} in the stratosphere.}
    \label{fig:all_modearth}
\end{figure*}

In this section, we present the results from our photochemical modeling with \Atmos for the baseline (i.e., observed MUSCLES spectrum) case and zero-continuum UV reconstructions.   
In general, we find that replacing the UV input with a reconstructed spectrum changes the abundances of photochemically active species. The differences prove significant, especially for our hazy (Archean) models, which exhibit the greatest deviation from our baseline models.

\subsection{Modern Earth} \label{sec:modern earth photo results}

We first present our \Atmos model outputs for modern Earth conditions for each of the MUSCLES catalog M-stars (Figure~\ref{fig:all_modearth}).  These models serve as our baseline case against which we will compare all of our UV-reconstructed models, and they also serve as a benchmark for comparison against similar previous works.  For example, \citet{Wunderlich2019} also modeled Earth-like planets orbiting the MUSCLES M-dwarf host stars  using a similar version of the \Atmos photochemistry-climate code but focusing on detectability of specific atmospheric spectral features.  

Overall, our models are in good agreement with \citet{Wunderlich2019}, with minor discrepancies being attributable to differences in model setup between our study and theirs.  For example, in \citet{Wunderlich2019}, to preserve ``Earth-like'' conditions, the authors ran their models varying the instellation such that the planetary surface retained the temperature of modern Earth's surface; whereas in our own work we retain Earth-like instellation across all of our models.  This choice leads to surface temperatures that are on average $\sim25$~K higher in our models compared to the fixed surface temperatures of \citet{Wunderlich2019}.  As a result of the different treatment of instellation, and also presumably due to other subtle differences in model implementation (e.g.\ reaction rates, opacities, etc.), the \citet{Wunderlich2019} version of \Atmos' climate model produces slightly differing temperature-pressure profiles compared to ours, including noticeably weaker (but still apparent) thermal inversions for the later-type M-stars.

These changes to the thermal structure of the atmosphere also result in notable differences in mixing ratios throughout the atmospheres. \ce{H2O}, which is parameterized below the tropopause as described in \revision{Section \ref{sec:atmos}}{\cite{Manabe1967}}, is directly tied to the thermal structure at these altitudes. Furthermore, the column depths of photochemical species such as \ce{O3} differ from the \citet{Wunderlich2019} models due to the temperature-sensitivity of their formation conditions.

Overall though, we achieve good qualitative agreement with trends seen in the \citet{Wunderlich2019} models as a function of spectral type.  Both of our works find more elevated upper-atmosphere temperatures above the tropopause with later spectral type and increasingly apparent stratospheric thermal inversions for later-type host stars.  We also both identify general trends of increasing H$_2$O, CH$_4$, CO, and N$_2$O column depths with later spectral type, accompanied by decreasing O$_3$.  

We find that our \Atmos models run with the reconstructed (zero-continuum) UV spectra compare favorably to our baseline models generated from the observed MUSCLES spectra (dashed lines vs.\ solid lines in Figure~\ref{fig:all_modearth}).  Some variations, particularly in species dominated by photochemistry such as \ce{O3}, have differences in column depth up to a factor of two to three. Some of the specific host stars, such as GJ 551 (Proxima Centauri), which have observed line luminosities that vary significantly from those calculated with the \citet{Melbourne2020} scaling relations, result in \Atmos model outputs that deviate more severely. We find overall though that reconstructed input UV spectra capture the bulk characteristics of these modern Earth-like atmospheres, and they therefore serve as suitable input to photochemical models in place of observations. For focused studies, variation in trace species or photochemically-dominated species may be significant enough to warrant a more careful treatment.

\subsection{Archean Earth} \label{sec:archean earth photo results}

\begin{figure*}[t]
    \centering
    \includegraphics[width=\textwidth]{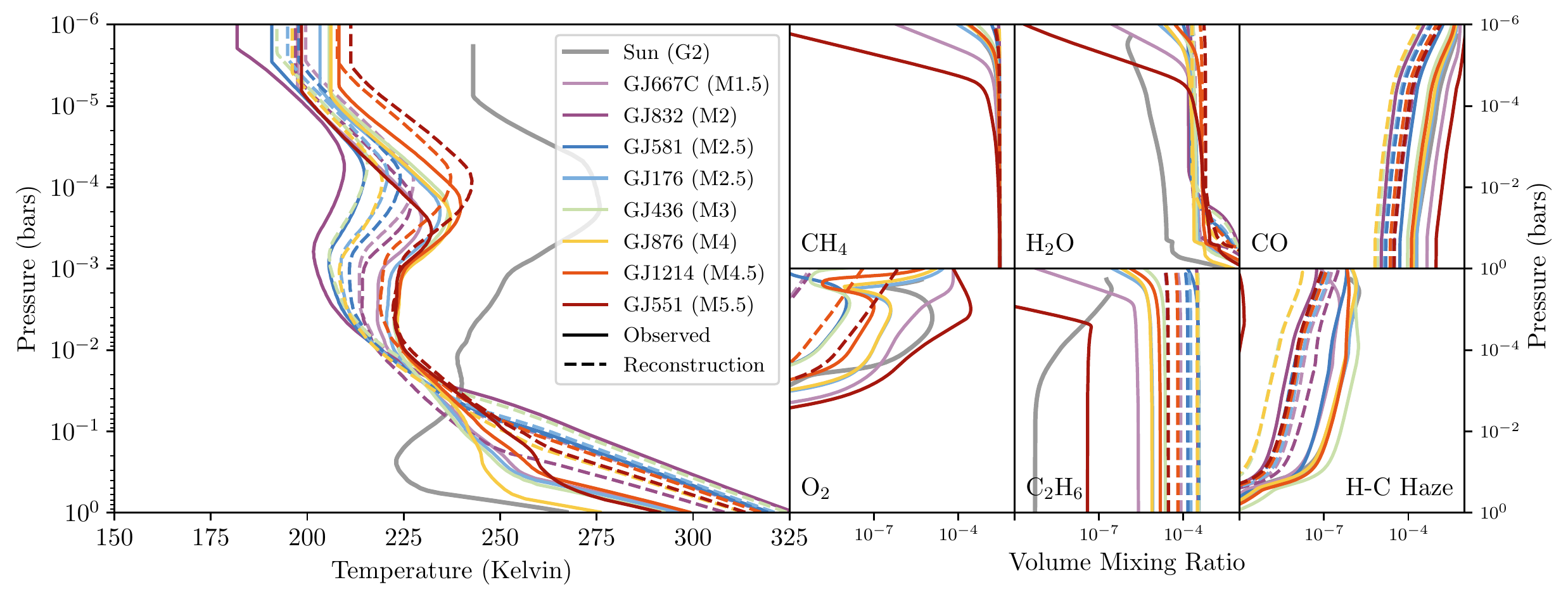}
    \caption{Same as Figure~\ref{fig:all_modearth}, but for our Archean Earth models including hydrocarbon haze.  Significant hydrocarbon haze ``precursors'' are generated in these models, e.g. \ce{C2H6}, as seen in the bottom middle mixing ratio panel.}
    \label{fig:all_archean}
\end{figure*}

    We similarly model a set of baseline \Atmos simulations for hazy, Archean Earth-like atmospheres (Figure~\ref{fig:all_archean}). These atmospheres have far more photochemically active species and the potential for strong radiative feedback from haze formation \citep{Arney2016}. This results in a greater sensitivity of the Archean Earth models to the UV input spectrum, compared to our modern Earth simulations.  
    
\begin{figure}[t]
    \centering
    \includegraphics[width=3.3in]{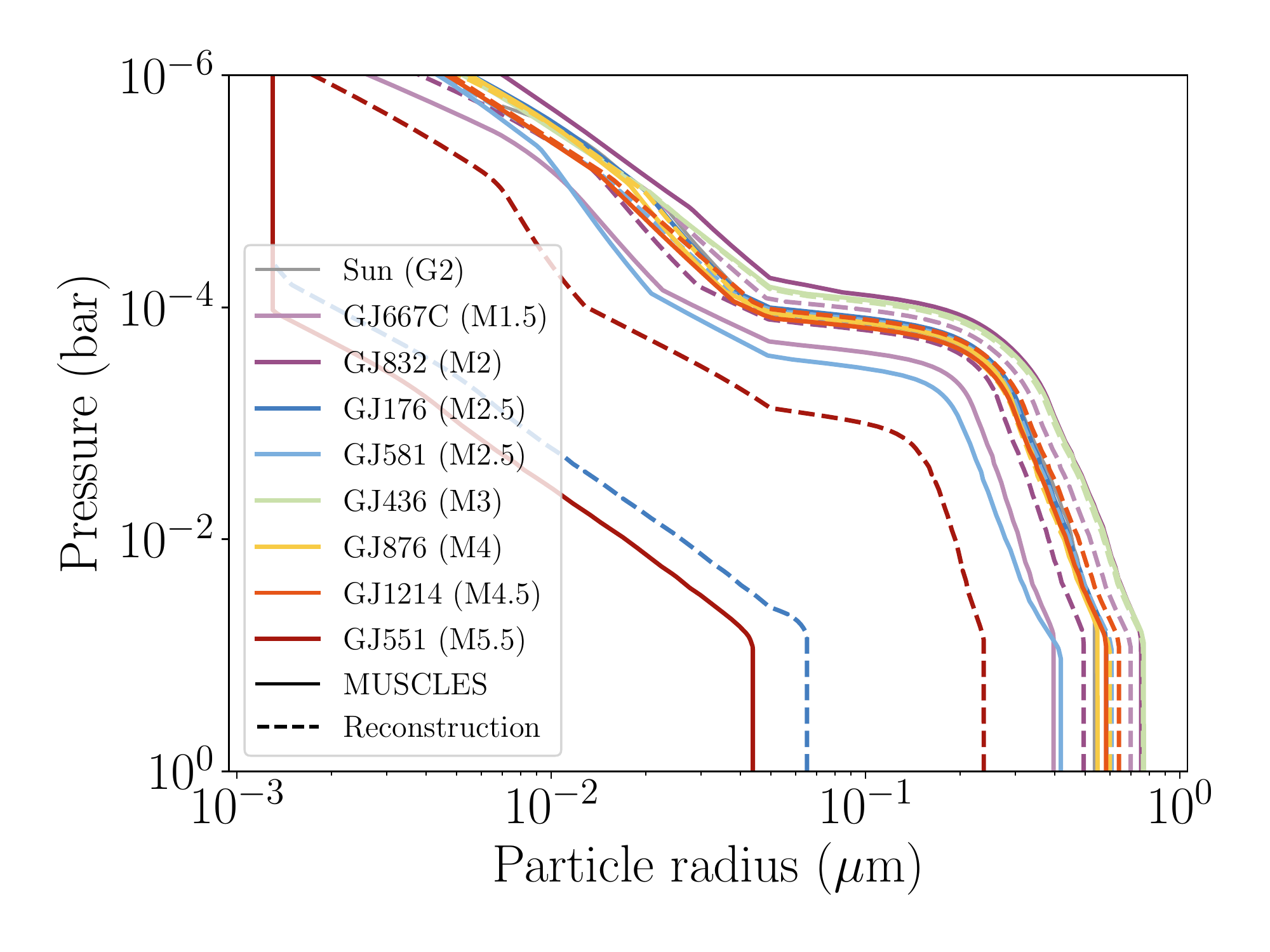}
    \caption{Vertical distribution of haze spherical radius. In general, higher incident UV fluxes produce haze particles that are able to sediment into larger radii at all altitudes. This is driven by higher abundances of hazes and haze precursors, which are able to agglomerate and sediment in the presence of more available haze particles. GJ 551 shows significantly smaller haze particle radii due to low abundances of haze precursors to polymerize into particles.}
    \label{fig:haze radius results}
\end{figure}

    Vertical abundance profiles for hydrocarbon hazes are shown in the lower right-hand panel of Figure~\ref{fig:all_archean}, and the corresponding particle size distributions in Figure~\ref{fig:haze radius results}.  The disparate optical properties for each \Atmos simulation (i.e., from the differing particle size distributions) couple with the thermal properties at haze-bearing altitudes to impact the chemical profiles of other species. These differences are non-linear, with feedback between the climate and photochemistry models playing significant roles in all characteristics of these more complex, hazy atmospheres. 
    
Compared to the modern Earth models, there are not such clear trends with stellar spectral type for both thermal structure and abundance profiles.  GJ 551 --- the latest M-star modeled --- does tend to be an end-member, but overall the lack of clear trends with stellar effective temperature imply that these hazy models are more sensitive to the activity level and exact details of the stellar UV spectrum than for the more ``well-behaved'' modern Earth models.  Of note, this increased sensitivity to the stellar UV is accompanied by increased challenges with model convergence --- it generally takes significantly longer for the hazy Archean Earth models to fully converge, and we typically must resort to the model ``short-stepping" procedure outlined in Section~\ref{sec:atmos}. 

The baseline models demonstrate the sensitivity of haze formation and feedback to small changes in the UV irradiation, and this is further shown when we use our reconstructed UV spectra as inputs. The reconstructed UV spectra often do not accurately reproduce the abundances of certain major species such as \ce{CO}, \ce{O2}, and \ce{C2H6} (shown in Figure \ref{fig:all_archean}). As for haze formation, in the majority of models, the abundances of haze particles are up to several orders of magnitude lower for our reconstructed spectra, resulting in a significant loss of UV opacity and changes to the thermal structure of the atmospheres.

An exception to our baseline models producing significantly more haze than the reconstructed models, GJ 551 produces orders of magnitude less haze using the MUSCLES spectrum as input compared to the reconstructed UV spectrum. This arises from the substantially higher line and continuum UV flux exhibited by GJ 551 compared to the other baseline input spectrum cases. This high irradiation photolyzes haze precursors that would otherwise polymerize into haze particles in the model. Coupled with oxidation of haze precursors due to photolysis of \ce{CO2} into oxygen radicals, the haze formation rate in the baseline GJ 551 model is negligible compared to the reconstruction case.

The degree of disagreement between the baseline and reconstructed models, and especially the systematic discrepancies in haze formation, leads us to suspect that other portions of the UV input spectra, beyond just the reconstructed emission lines, may be playing an important role.  We revisit this idea in depth in Section~\ref{sec:continuum treatment results}.  In the meantime, we conclude that the \citet{Melbourne2020} UV reconstructions may not adequate for modeling the photochemistry in hydrocarbon haze-producing atmospheres.

\section{Transmission spectra results} \label{sec:transmission spectra results}

We next examine how our model-derived temperature and abundance profiles impact the observable properties of the simulated exoplanets --- specifically their transmission spectra between 0.3 and 30 $\mu$m.

\subsection{Modern Earth} \label{sec:modern earth trans results}

\begin{figure*}[t]
    \centering
    \includegraphics[width=\textwidth]{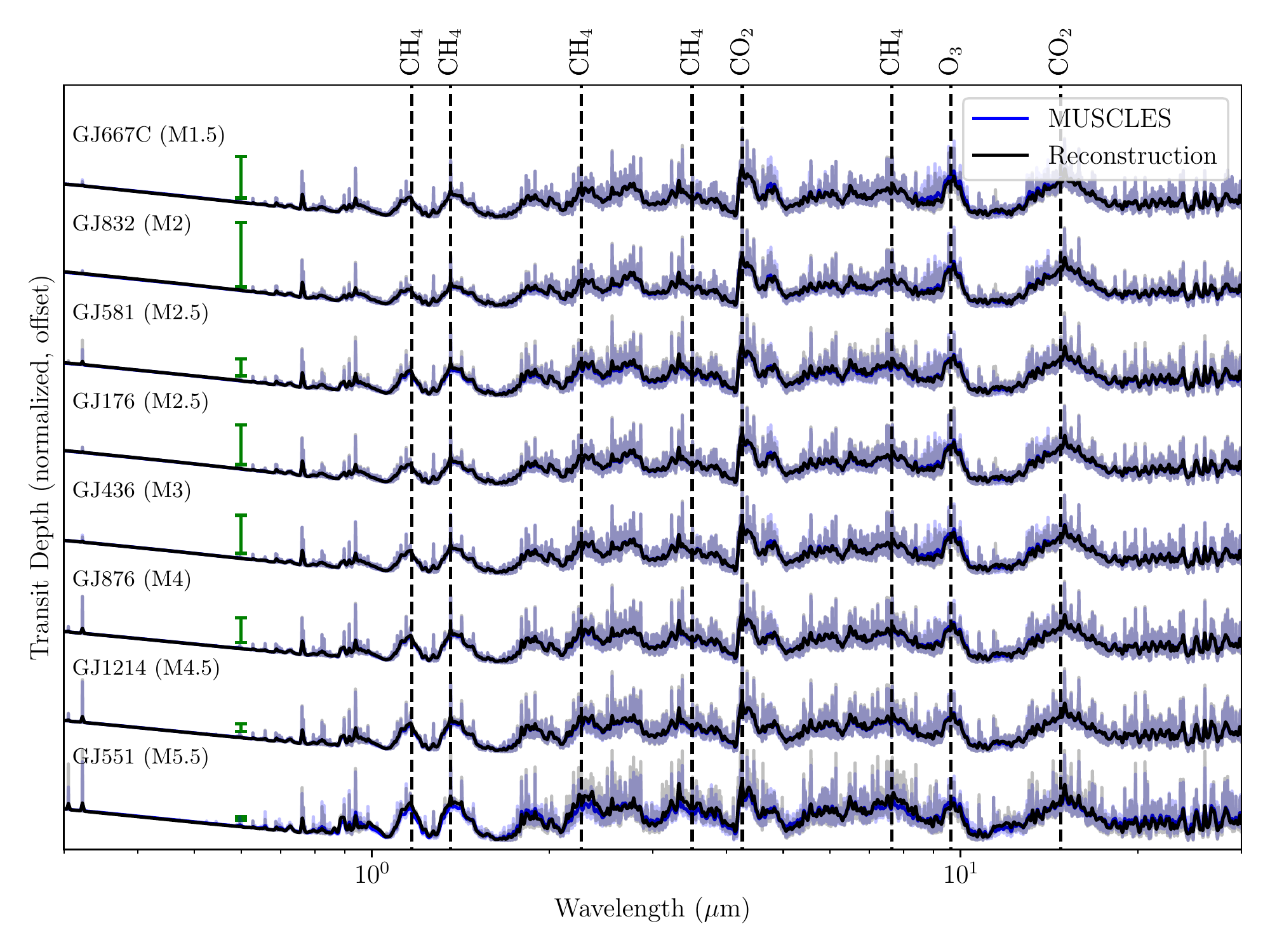}
    \caption{Model transmission spectra of our Earth-like atmospheres. The black and blue curves are our baseline models with the raw MUSCLES spectrum and reconstructed (zero continuum) UV spectrum as inputs, respectively. The lighter colored lines are the full-resolution transmission spectra output by \exotransmit (at a spectral resolution of $R = 1000$), while the dark, thick lines are smoothed for ease of visualization. Transit depths in both cases are normalized to the transmission spectrum for the baseline (MUSCLES) model. For reference, the green bars indicate an amplitude of 5 ppm for each host star, assuming the transiting exoplanet is Earth-size.  For larger and/or hotter planets, these error bars will shrink proportionately.  Spectral features of key molecules are indicated.}
    \label{fig:all_modearth_transmission_spectra}
\end{figure*}

As already shown in Figure~\ref{fig:all_modearth}, the dominant absorbing species in the modern Earth models do not differ significantly between our reconstructed and baseline scenarios, especially at the $\sim$mbar pressures probed by transmission spectroscopy. As a result, our \texttt{Exo-Transmit} transmission spectrum models  (Figure \ref{fig:all_modearth_transmission_spectra}), show negligible differences in the wavelength-dependent absorption produced by any of these model atmospheres, when comparing the reconstructed to baseline cases. 

For this set of modern Earth models, the largest absolute change in transit depth encountered between baseline and reconstructed models is 5~ppm (indicated by green bars for each of the spectra in Figure \ref{fig:all_modearth_transmission_spectra}), which is below the anticipated noise floor for observatories like JWST.   We note that while the 5~ppm bars in Figure~\ref{fig:all_modearth_transmission_spectra} also make it look as though many of the modeled atmospheres are well out of reach for atmospheric characterization with JWST \citep[with an expected noise floor for many instruments at the $\sim$10--20 ppm level;][]{taro19,schlawin20,schlawin21}, these bars will scale down proportionally to both the planet's radius and its equilibrium temperature, indicating that, all else being equal, hotter and larger planets are easier to characterize \revision{}{(further discussed in Section \ref{sec:archean earth trans results})}. \revision{}{That said, small, Earth-like planets will be characterizable with JWST with sufficient integration time, such as those of TRAPPIST-1 and some super-Earth TESS discoveries within the habitable zone \citep[e.g.,][]{Deming2009,Molliere2017,Meadows2018}.}

Our main finding here is that prominent broadband absorbers such as \ce{H2O}, \ce{CO2}, \ce{CO}, and \ce{CH4} --- and therefore the transmission spectra themselves --- are essentially insensitive to UV spectrum reconstruction for a modern Earth atmospheric scenario.

\subsection{Archean Earth} \label{sec:archean earth trans results}

\begin{figure*}[t]
    \centering
    \includegraphics[width=\textwidth]{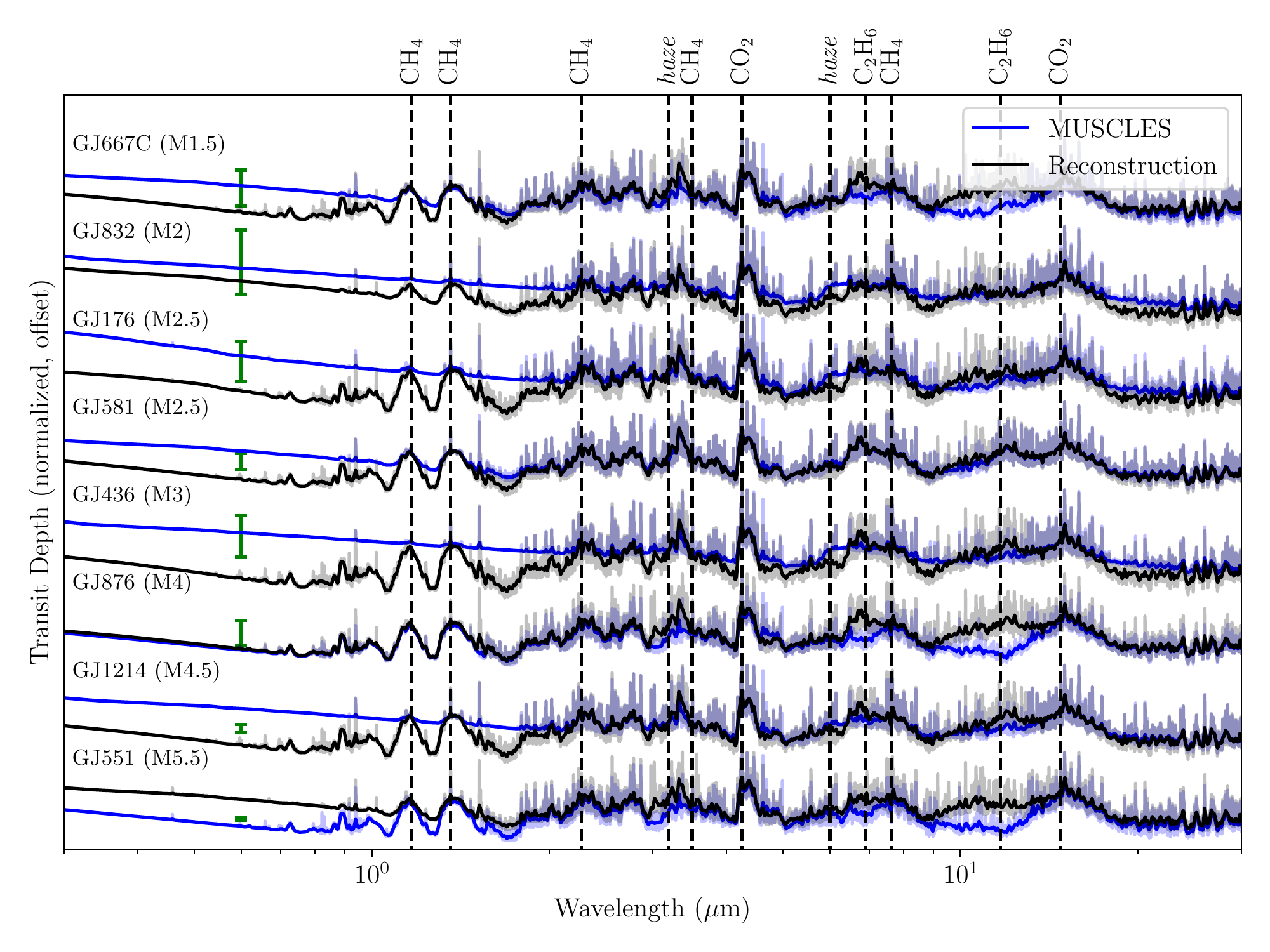}
    \caption{Same as Figure~\ref{fig:all_modearth_transmission_spectra} but for the Archean Earth models. '\textit{haze}' indicates features associated with haze particles. For all stars with the exception of GJ 551, the model produces significantly less haze using the reconstructed UV spectra. In the case of GJ 551, our reconstruction actually produces more haze than in the baseline case (see Section \ref{sec:archean earth photo results}.)}
    \label{fig:all_archean_transmission_spectra}
\end{figure*}

The Archean models behave more dramatically with respect to UV reconstructions.  Figure \ref{fig:all_archean_transmission_spectra} demonstrates the significant variations between hazy models produced in the baseline vs.\ reconstructed scenarios. The most significant differences between pairs of transmission spectra comes from the haze opacity itself, seen as a gentle downward slope, most prominent in the optical and UV.  As discussed in Section~\ref{sec:archean earth photo results}, the reconstructed models generally produce less haze than our baseline models, resulting in shallower transit depths and more prominent molecular absorption features at wavelengths shorter than 3 $\mu$m.  Differences in the optical scattering slopes between spectra are related to the differing particle size and vertical haze distributions.  The magnitude of the differences between baseline and reconstructed transmission spectra also varies considerably with host star, but with no clear progression as a function of spectral type, in agreement with our photochemical modeling results from Section~\ref{sec:archean earth photo results}.

Significant discrepancies of up to 20 ppm also arise between our hazy model transmission spectra at IR wavelengths of $\sim$6, 7, 12, and 20 $\mu$m, as seen in Figure~\ref{fig:all_archean_transmission_spectra}. Differences at 7 and 12 $\mu$m are attributable to variations in the amount of \ce{C2H6} across atmospheres, which is not as efficiently photolyzed by the reconstructed UV spectra. Differences at 6 and 20 $\mu$m are a result of differences in haze abundance and optical properties.

The $\sim$20-ppm differences between our baseline and UV reconstructed models are expected to be marginally distinguishable by JWST, and therefore using reconstructed UV spectra will potentially have observable consequences for these Archean Earth atmospheres.  Furthermore, larger and/or hotter planets will produce even larger transmission spectral features, leading to more obvious differences between baseline and reconstructed scenarios for hazy atmospheres\revision{, all else being equal}{. More massive planets are more likely to host hydrogen-dominated atmospheres, resulting in significantly different chemical networks and haze formation pathways when compared to terrestrial atmospheres \citep[e.g.,][]{He_2020}. As a result, differences in bulk composition may impact the nature of hazes produced \citep{Moran_2020}, as well as their vertical distribution throughout the atmosphere, when compared directly to the terrestrial models we use in this study.}  

\section{UV Continuum treatment results} \label{sec:continuum treatment results}

\subsection{Continuum models}\label{sec:continuum modeling results.}

\begin{figure*}[t]
    \centering
    \includegraphics[width=\textwidth]{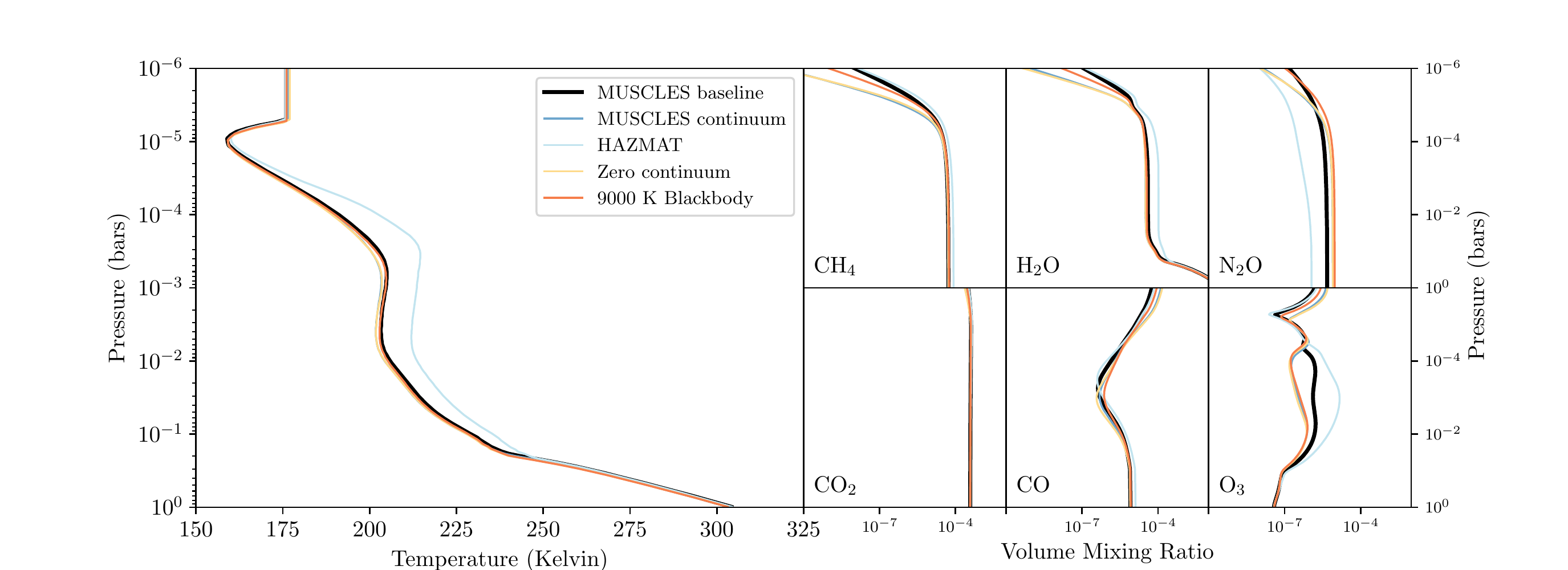}
    \caption{Temperature-pressure profiles and abundance profiles for different UV continuum treatments for our modern Earth-like models, specifically using the GJ 176 input case. ``MUSCLES baseline'' is the model using the raw (observed) MUSCLES UV spectrum. The rest of the models shown use the \citet{Melbourne2020} scaling relations to reproduce the star's UV emission lines, and one of our continuum reconstructions, as indicated. The majority of continuum treatments reproduce the baseline model, and no observationally significant differences arise across models (see Figure~\ref{fig:gj176 archean spectra}), though photochemically sensitive species such as \ce{O3} and \ce{N2O} deviate for certain continuum treatments.}
    \label{fig:gj176 modern profiles}
\end{figure*}

\begin{figure*}[t]
    \centering
    \includegraphics[width=\textwidth]{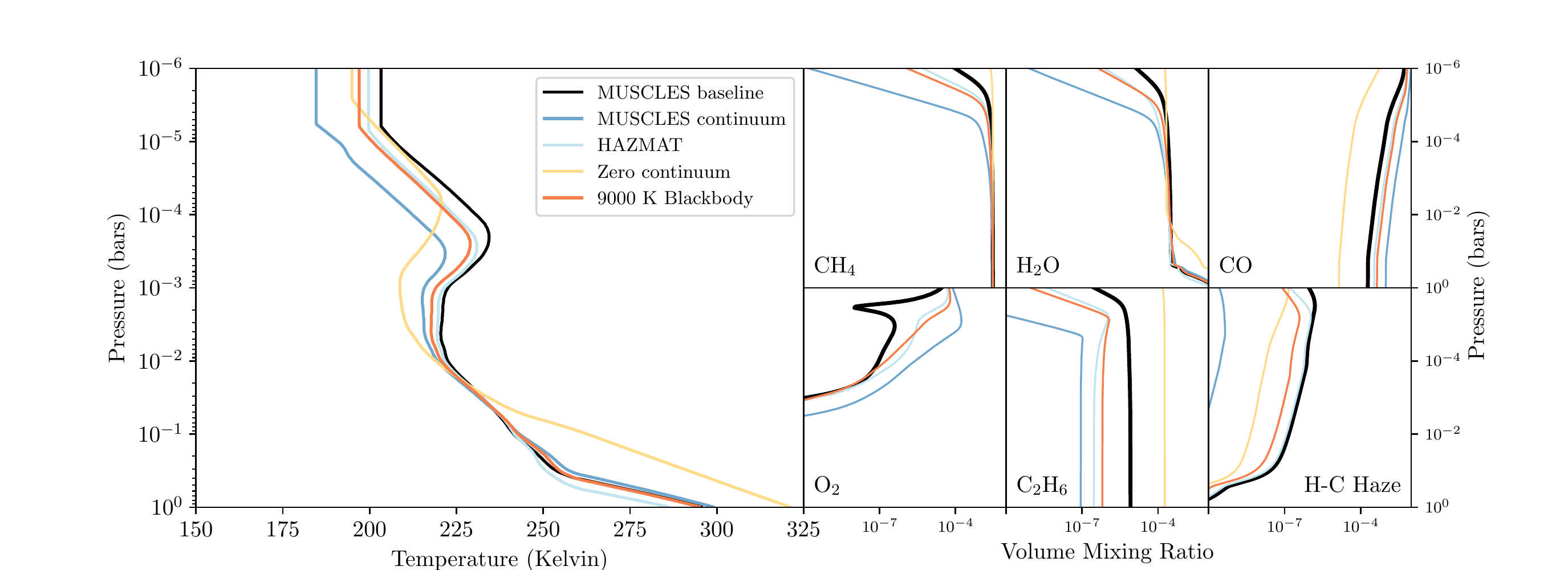}
    \caption{Same as Figure~\ref{fig:gj176 modern profiles}, but for our hazy Archean Earth-like models. These haze-forming models are much more sensitive to the properties of the host star's UV continuum, particularly for hydrocarbon hazes and their precursors. These lead to observable differences, shown in Figure~\ref{fig:gj176 archean spectra}.}
    \label{fig:gj176 archean profiles}
\end{figure*}

\begin{figure*}[t]
    \centering
    \includegraphics[width=\textwidth]{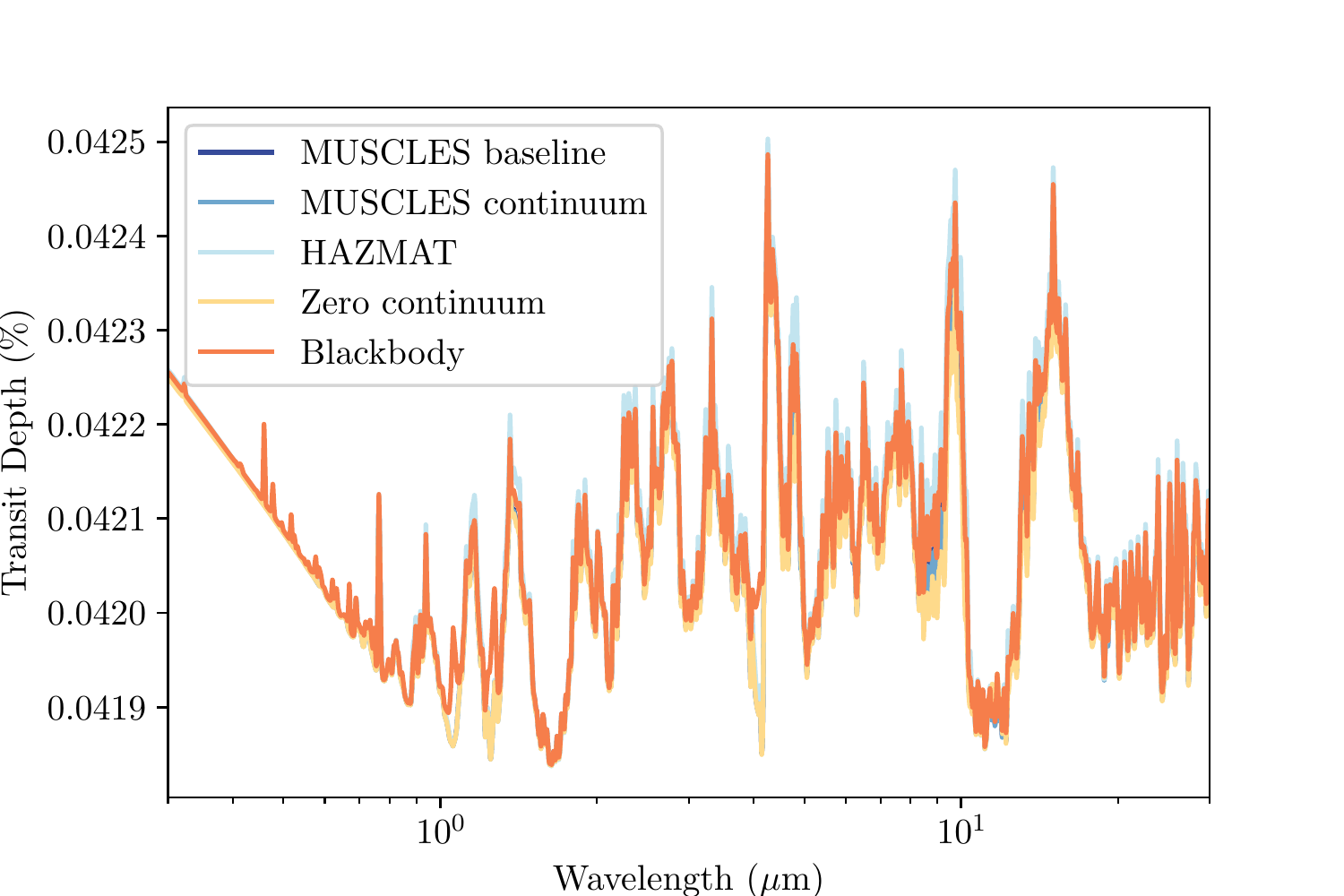}
    \caption{Transmission spectra for modern Earth models of GJ 176 using different prescriptions for reconstructing the host star's UV continuum, as indicated. The maximum difference of 2 ppm occurs at 1.5 $\mu$m, a \ce{CH4} line, between the baseline and no-continuum case.}
    \label{fig:gj176 modern spectra}
\end{figure*}

\begin{figure*}[t]
    \centering
    \includegraphics[width=\textwidth]{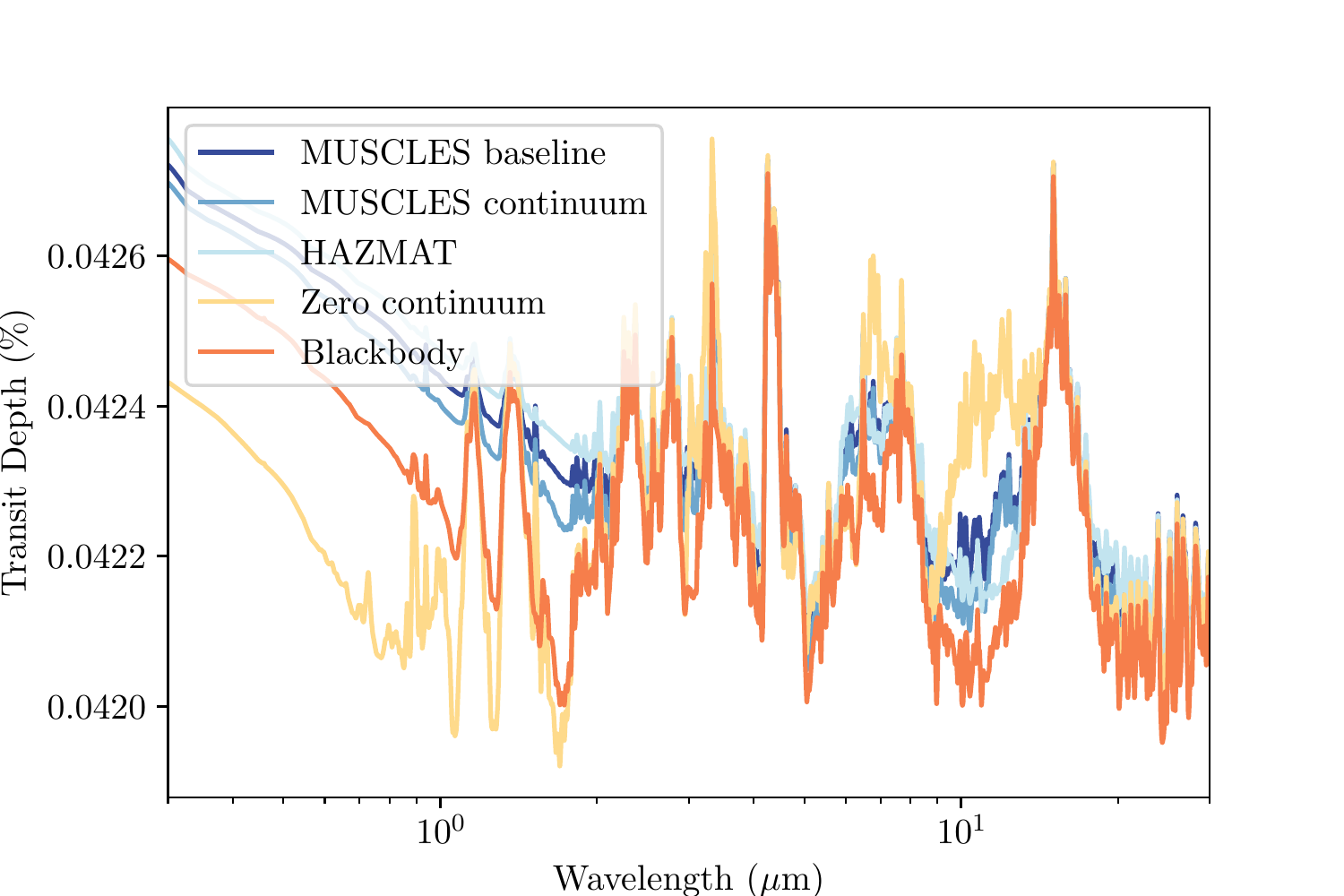}
    \caption{Same as Figure~\ref{fig:gj176 modern spectra}, using hazy Archean Earth-like models. A maximum difference between the baseline MUSCLES spectrum and the continuum treatments of 6 ppm occurs at 1.01$\mu$m for the zero-continuum case.}
    \label{fig:gj176 archean spectra}
\end{figure*}

\begin{figure}[t]
    \centering
    \includegraphics[width=3.3in]{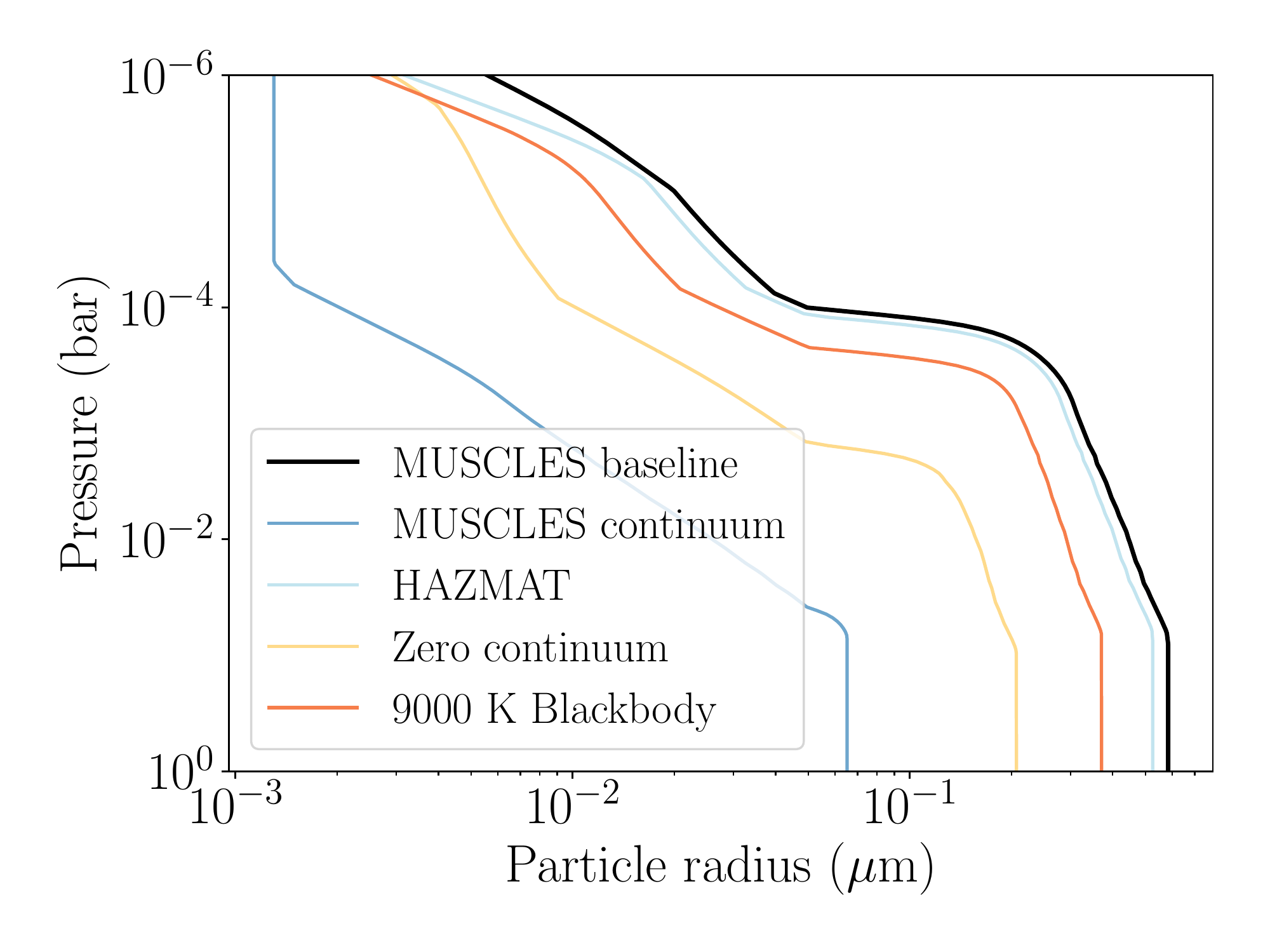}
    \caption{\revision{}{Vertical distribution of compact sphere equivalent radius for the various continuum treatments. All reconstructions produce smaller haze particles compared to our GJ 176 MUSCLES baseline case. In general, particle size varies with the total integrated continuum flux. However, when using the MUSCLES continuum flux, very small particles are created compared to any other reconstruction treatments. This is due to significantly variable photolysis rates, reported in the relevant supplementary data for each of our cases }.}
    
    \label{fig:haze radius results continuum}
\end{figure}

Figures \ref{fig:gj176 modern profiles} and \ref{fig:gj176 archean profiles} show the results of using different UV continuum treatments, described in Section~\ref{sec:MUSCLES}, for our modern and Archean Earth-like models of GJ 176, respectively.  In the hazeless modern Earth atmospheres, the choice of continuum treatment has an observationally insignificant impact on the resulting abundance and temperature profiles.  The abundances of the most prominent photoactive species (e.g.\ \ce{O3}, \ce{N2O}) are found to moderately depend on the magnitude of the continuum flux.  Notably, the \ce{O3} abundance profile deviates significantly for all continuum treatments. These deviations do not prove observationally significant, as shown in Figure~\ref{fig:gj176 modern spectra}. 

Our hazy Archean Earth-like models prove more sensitive to choice of UV continuum. Figure~\ref{fig:gj176 archean profiles} demonstrates the severity of these discrepancies, with the worst case scenarios producing several orders of magnitude less haze than models using the baseline MUSCLES UV input spectrum. This translates to molecular species such as \ce{CO}, \ce{O2}, and haze precursors like \ce{C2H6} differing significantly as well.

As a result of the sensitivity to the UV continuum in the Archean models, the model transmission spectra (Figure~\ref{fig:gj176 archean spectra}) also differ substantially.  This is especially true in the visible and near-IR where the haze impacts on the transmission spectra are most apparent.  A comparison between the reconstructed model with zero continuum vs.\ the one with the observed MUSCLES continuum (blue vs.\ orange line in Figure~\ref{fig:gj176 archean spectra}) is especially telling of the role that the UV continuum plays in shaping the properties of hazy atmospheres.  Ignoring the continuum entirely clearly neglects an important haze formation pathway and also impacts the transmission spectra via changes to thermal structures and the abundances of other key absorbers.  

The HAZMAT reconstruction best reproduces the transmission spectrum generated from the MUSCLES observations for GJ 176, although the somewhat higher UV continuum in this reconstruction (see Figure~\ref{fig:gj176 all continua}) leads to a modest over-production of haze (seen as a deeper optical transit depth). Overall, we conclude that the UV continuum treatment definitely plays a non-negligible role when modeling hazy atmospheres.

\subsection{Using a neighboring stellar type} \label{sec:neighbor type results}

Another way to account for continuum flux is to assume that the continuum of an observed star similar in stellar type will be sufficient for UV reconstruction. To that end, we provide a brief assessment of the applicability of such an approach in our GJ 176 reconstruction case. 

Rather than using a \emph{model} of the UV continuum, employing \emph{observed} UV data for an actual star may allow us to account for physical effects not adequately captured by a model spectrum.  Furthermore, one could hypothetically attempt to match a proxy star's observed continuum to other physical properties of the star one is reconstructing. For example, if a star with similar activity has an observed UV spectrum it may be better than a model or a quiescent spectrum for a neighboring stellar type.

\begin{figure*}[t]
    \centering
    \includegraphics[width=6in]{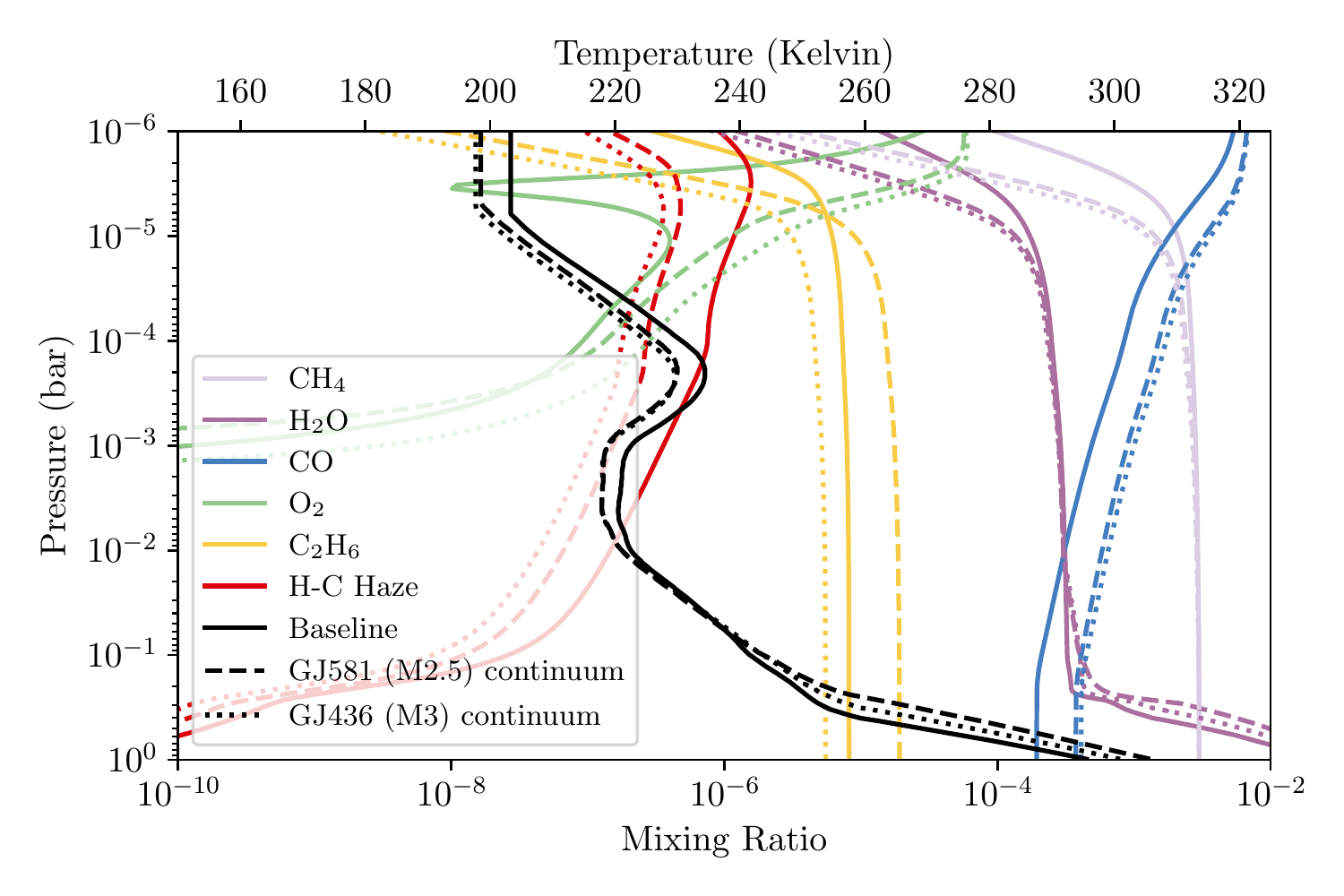}
    \caption{Model results using two MUSCLES stars of neighboring spectral types as proxy continua for a GJ176 reconstructed spectrum. Carbon-bearing species such as \ce{CO}, \ce{CH4}, and \ce{C2H6} all have significantly different abundances compared to the baseline case, with the haze nearly an order of magnitude less abundant in either case.}
    \label{fig:neighboring star type treatment}
\end{figure*}

Figure~\ref{fig:neighboring star type treatment} shows the results of using the GJ 176 line reconstructions, as done previously (Figures \ref{fig:gj176 modern profiles} and \ref{fig:gj176 archean profiles}), but using two different MUSCLES stars for the UV continuum fluxes. In this case, we choose the two stars closest in stellar type compared to GJ 176 (an M2.5V star): GJ 581 (M2.5V) and GJ 436 (M3V). Here we only model Archean Earth conditions --- as Sections \ref{sec:modern earth photo results} and \ref{sec:modern earth trans results} have already shown the relative insensitivity of our hazeless (modern Earth) models to the UV continuum. We reconstruct GJ 176's UV spectrum using one of the neighboring star's UV continuum to fill between our reconstructed lines using the GJ 551 $R'_{HK}$ values. At non-UV wavelengths the MUSCLES GJ 176 panchromatic spectrum is used (relevant only for the climate model). Once the spectrum reconstruction is complete, the full spectrum is re-normalized to have a total integrated instellation equivalent to Earth-equivalent flux at the top of the atmosphere.

We find that the continua of both stars do a reasonable job of replicating the baseline GJ 176 model, but again noticeable differences do arise.  Haze abundances between the three models differ by up to an order of magnitude, and abundances of various molecules (e.g.\ \ce{CO}, \ce{CH4}, \ce{C2H6}, \ce{O2}) similarly disagree by factors of a few, as seen in Figure~\ref{fig:neighboring star type treatment}.

\begin{figure*}
    \centering
    \includegraphics[width=6in]{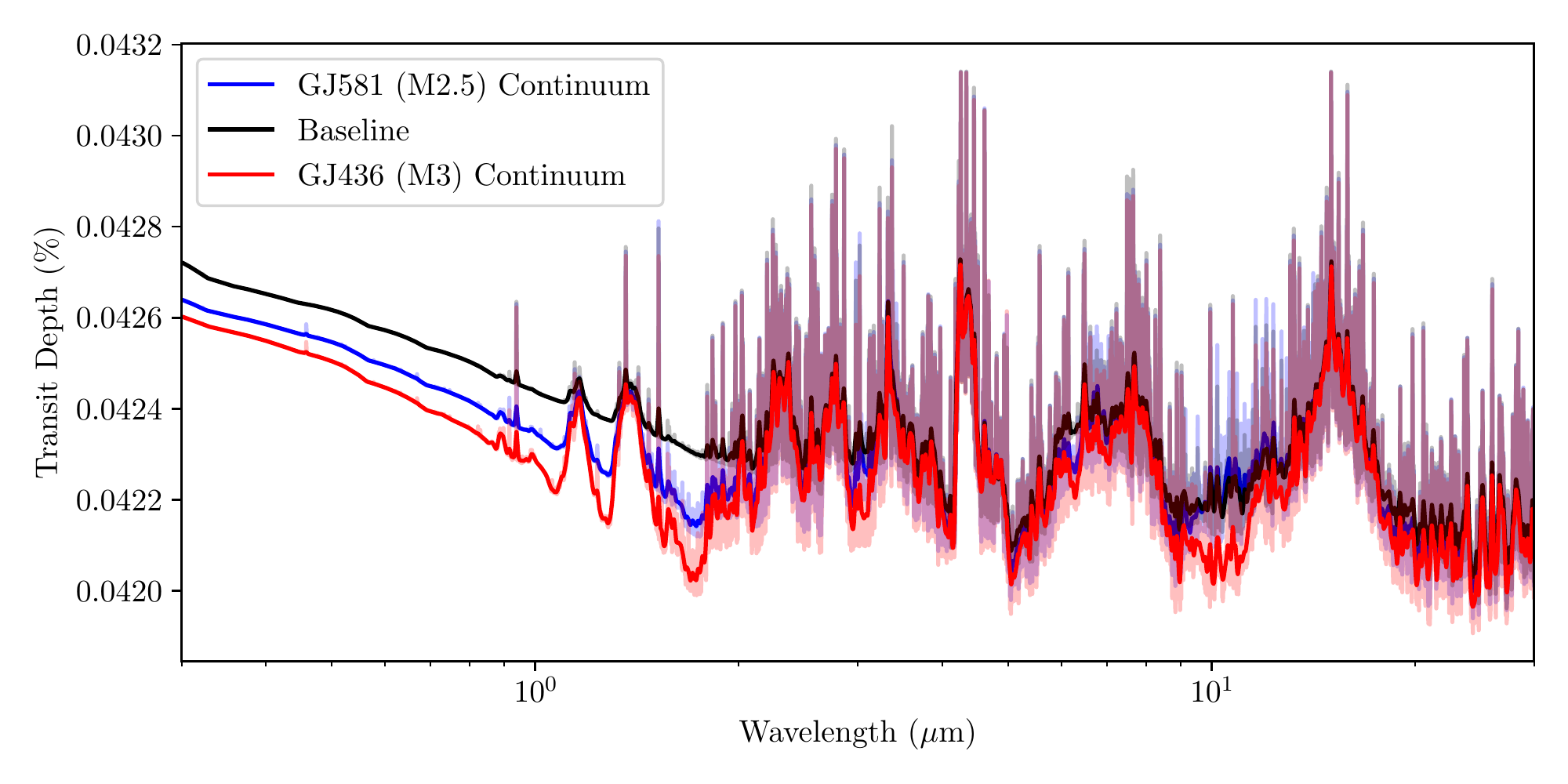}
    \caption{Transmission spectra resulting from the model results in Figure~\ref{fig:neighboring star type treatment}. The different haze abundances shown there are reflected in the transmission spectra's haze absorption at wavelengths shortward of $\sim 2$ $\mu$m. Furthermore, a feature at 10~$\mu$m is caused by different \ce{C2H6} abundances in the model using the GJ 436 continuum as a proxy.}
    \label{fig:neighboring star type transmission spectra results}
\end{figure*}

As for the resulting transmission spectra (Figure~\ref{fig:neighboring star type transmission spectra results}), significant discrepancies again arise at optical wavelengths due to differing haze abundances and particle sizes among the three models, and also at $\sim 12$ $\mu$m from \ce{C2H6}, \ce{C2H4}, and \ce{C2H2} absorption.  Interestingly, the M2.5 star GJ 581 does manage to replicate the baseline GJ 176 model (itself an M2.5 star) with reasonable accuracy, but only longward of 2 $\mu$m.  At shorter wavelengths the differences in haze properties become apparent.  These differences in simulated transmission spectra for two stars of identical spectral classification indicate that spectral type is not a unique predictor of photochemical behavior of an exoplanetary atmosphere, nor of its observable properties.

\begin{figure}[t]
    \centering
    \includegraphics[width=3.3in]{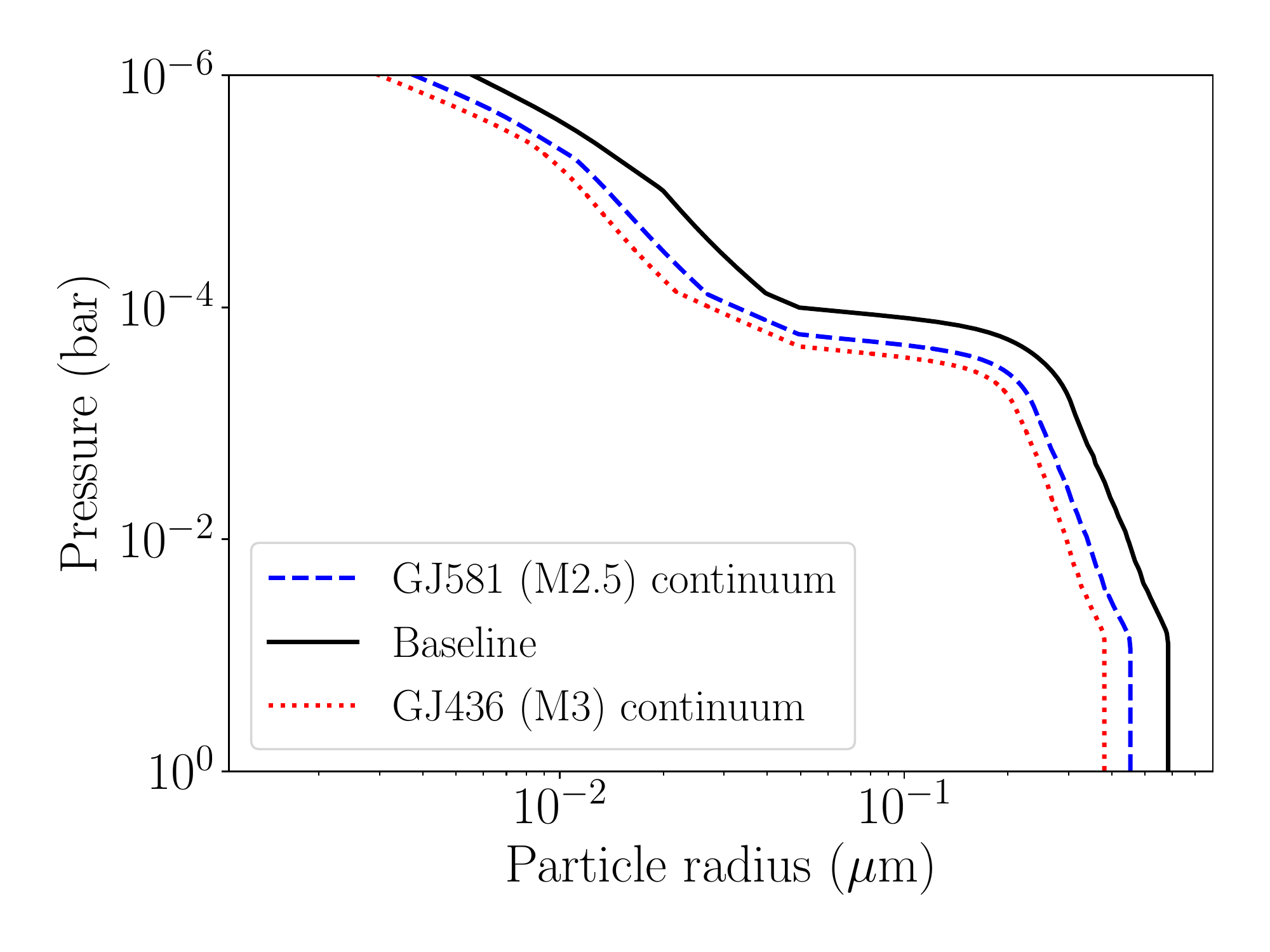}
    \caption{\revision{}{Vertical distribution of haze spherical radius using continuum from MUSCLES stars of neighboring stellar types compared to GJ 176 (of type M2.5), as in Figures \ref{fig:neighboring star type treatment} and \ref{fig:neighboring star type transmission spectra results}.} 
    }
    \label{fig:haze radius results neighbors}
\end{figure}

\subsection{Which UV continuum treatment is correct?}
\label{sec:correct continuum treatment}

Based on our results, models with photochemical hazes can vary dramatically, even at an observable level, as a function of the UV continuum treatment applied.  With this in mind, it is important to establish which UV continuum treatment is the ``best" one to use in the absence of UV observations.  

From Figures~\ref{fig:gj176 archean spectra} and \ref{fig:neighboring star type transmission spectra results}, we see that the HAZMAT and neighboring host star continuum treatments do the best job of replicating the behavior of our baseline Archean Earth model for GJ 176 from an observational perspective.  In practice, semi-empirical spectral models have only been generated so far for a very limited number of host stars \citep{Fontenla2016, Peacock2019a, Peacock2019b, Tilipman2021}, GJ 176 being one of them, and rely on observations of the host stars' UV spectra for their models. As a result, in many cases it may be impractical to use such a model for replicating the UV continuum of an arbitrary exoplanet host star.  Similarly, using observations of a neighboring spectral type for UV continuum reconstruction is also problematic, for reasons discussed in Section~\ref{sec:neighbor type results} --- mainly that spectral type is not a unique predictor of UV continuum behavior. \revision{}{This also applies to using continuum flux from a star of the same stellar type, since factors such as age, composition, and activity level will all affect the strength of the continuum for a given host star.} However, in many situations this may be the most practical solution, especially as the library of observed M star UV spectra continues to grow. It should be noted though that the ``neighboring spectral type" approach is not guaranteed to produce accurate outcomes for photochemical modeling.

Two further caveats to this finding are as follows.  First of all, we have only done a detailed examination of the UV continuum treatment for a single host star --- GJ 176 --- and therefore our results may not be fully generalizable.  Secondly, we have based our modeling approach off of the premise that the observed MUSCLES spectrum represents the ground truth of the host star's UV output.  Unfortunately, due to the intrinsic UV-faintness of many M dwarfs, the MUSCLES-reported continuum fluxes are often representative of the photon-limited noise floor of of the data, rather than a true detection of the stellar emission \citep{Loyd2016}.  As a result, the MUSCLES UV continuum fluxes may be overestimates of the true UV emission, especially for fainter stars and at shorter wavelengths.  

\section{Conclusions} \label{sec:conclusion}

To summarize our study and its findings:
\begin{enumerate}
    \item We have used the MUSCLES Treasury survey M-dwarf spectra \citep{France2016} coupled with UV reconstructions from \citet{Melbourne2020} to generate photochemical and transmission spectrum models of terrestrial atmospheres at Earth-like instellation, with and without hazes.
    \item We find these reconstructions to be adequate for photochemical modeling of hazeless (modern Earth-like) terrestrial atmospheres. Deviations from our baseline models (i.e., those using the MUSCLES observations) are minimal, though species formed primarily via photochemistry are slightly underestimated.  Models generated from MUSCLES observations and UV reconstructions produce nearly identical transmission spectra.
    \item Photochemical models of hazy (Archean Earth) terrestrial planets are much more sensitive to the UV input spectrum. Chemical abundances, haze formation rates, and thermal profiles are all significantly impacted by the use of UV reconstructions of host star spectra.  These changes to the atmospheric structure and chemistry have observable implications in the transmission spectra of hazy exoplanets. 
    \item We further find that our hazy atmosphere results are sensitive to the UV continuum flux, which is not modeled in our nominal UV reconstructions.  Changes in continuum fluxes --- tested on our models of the early M-star GJ 176 --- are also found to impact thermal structures, as well as haze and molecular abundances, at an observable level, resulting in transit spectra with significant differences across a broad range of wavelengths.
\end{enumerate}

Given these results, we find that fully observing a host star's UV spectrum, including multiple UV emission lines and the underlying continuum, remains the gold standard for modeling exoplanet atmospheres. While the \citet{Melbourne2020} reconstructions are a good proxy for M-dwarf stellar spectra for the purpose of photochemically modeling non-hazy Earth-like atmospheres, they do not sufficiently capture a star's UV spectrum for hazy planet modeling.  For that reason, the observed stellar spectrum is especially necessary for predicting and interpreting transmission spectra of hazy exoplanets.

For cases in which it is not possible to observe the host star's UV spectrum, we recommend the following procedure.  

\begin{itemize}
\item Reconstruct the strongest UV emission lines using the \citet{Melbourne2020} scaling relations.  This requires knowledge of star's the $R'_{HK}$ index as well as an estimate of its bolometric luminosity --- both of which should be readily obtainable through optical characterization.  If certain UV lines (e.g.\ Ly~$\alpha$) have been observed, but the rest of the UV spectrum has not, one can use those observed line fluxes in tandem with the \citet{Melbourne2020} scaling relations to fill in the fluxes of the remaining emission lines. \revision{}{We reiterate here that \cite{Melbourne2020} find $R'_{HK}$ to be the most robust predictor of UV emission line strength across the M-dwarf spectral class, and therefore UV emission lines---in the absence of direct observations---should be generated following the $R'_{HK}$ scaling relations.}

\item To reconstruct the UV continuum, either choose an observation of a star with a similar spectral type, as done in Section~\ref{sec:neighbor type results}, or employ a synthetic model of the UV continuum, such as those provided by the HAZMAT program \revision{Peacock et al. 2019b}{\citep{Peacock2019b}}. We have found that these two options produce model results most consistent with our baseline cases using the MUSCLES observed  UV spectra. For cases in which neither of these two approaches are feasible, a blackbody continuum can be used, following the example of \cite{Ayres1979}. 

\end{itemize}

We note that it remains problematic to make use of observations of the stellar UV continuum in photochemical modeling because in many cases those observations simply represent the photon noise level, rather than a true detection of the stellar emission.  We therefore recommend deeper observations of a benchmark set of exoplanet host stars that fully detect and resolve the UV continuum emission, accompanied by improved modeling of M-dwarf spectra in the UV. \revision{}{However, such observations may not be feasible for most exoplanet host stars without a more sensitive far-UV observatory, such as the 6-m UV/optical/IR observatory recommended by the Astro2020 Decadal Survey \citep{2020decadalsurvey}.}

In order to generate our photochemical models for well-benchmarked cases, the planetary scenarios we've studied have low effective temperatures and masses compared to many favorable targets for atmospheric characterization. The trends and conclusions reached in this study can likely be generalized to larger, warmer exoplanets, although for such planets disequilibrium processes are less dominant in establishing atmospheric composition. To more accurately predict trends in such atmospheres would require an extension of this study covering a broader parameter space. 

Stellar activity changes the time-averaged high-energy irradiation of a planet's atmosphere, altering the photochemical equilibrium of a planet's atmosphere depending on the rate of flaring \citep{Segura2010} and the stellar magnetic activity cycle. Some information about the activity for the MUSCLES target stars is folded into their observed spectra due to flares occurring during exposure time \citep{France2016,Loyd2018}, though this does not provide sufficient information to draw conclusions in our work. Since the UV flux during flaring events can increase by several orders of magnitude, understanding how these events change the time-dependent evolution of these atmospheres would improve upon the results, particularly in the case of haze-forming atmospheres.

As the community prepares for future space- and ground-based observatories capable of unprecedented atmospheric characterization, it is critical to understand what complementary data sets will be required to contextualize and interpret these future studies. Here, we have focused on the role that UV observations play in accurate modeling of disequilibrium chemistry in exoplanetary atmospheres.  Our results have implications for addressing compelling questions in astrobiology, atmospheric evolution, and aerosol formation --- all of which are fundamentally tied to the photochemistry occurring in a planet's atmosphere.  Our work motivates the use of the aging HST facility to perform UV observations of exoplanet host stars at high precision as a critical input to photochemical models.  Following the demise of HST, future UV missions from the flagship to the \revision{probe}{SmallSat} scale, \revision{}{such as the 6-m UV/optical/IR flagship recommended by the Astro2020 decadal to smaller observatories in the nearer term on the Explorer and SmallSat scales} like \revision{\textit{LUVOIR} (The LUVOIR Team 2019), \textit{HabEx} (Gaudi et al. 2019), \textit{CETUS} (Heap et al. 2019), \textit{ESCAPE} (France et al. 2020b), }{}\textit{CUTE} \citep{CUTE} and SPARCS \citep{SPARCS} will have an important role to play in providing further UV information for exoplanet host stars.  In the absence of UV observations, proxy scaling relations and UV reconstruction techniques remain the best path forward.

\acknowledgments
{
D.J.T. and E.M.-R.K. acknowledge funding from the NSF AAG program (grant \#2009095) and from the Hubble Space Telescope theory program (HST-AR-16135).  We thank E.T. Wolf for generating the \ce{H2O} and \ce{CO2} k-coefficient tables used in the climate model, as well as Nick Wogan for assistance updating photochemical cross sections in \Atmos. G.A. acknowledges support from the Virtual Planetary Laboratory, a member of the NASA Nexus for Exoplanet System Science (NExSS) research coordination network Grant 80NSSC18K0829. S.B. acknowledges support from NASA under award number 80GSFC21M0002. S.B., A.Y., and G.A. acknowledge support from the Goddard Space Flight Center Sellers Exoplanet Environments Collaboration (SEEC) for support, which is funded by the NASA Planetary Science Division’s Internal Scientist Funding Model (ISFM). 
}

\newpage

%


\software{
\href{https://www.astropy.org/index.html}{\texttt{astropy}} \citep{astropy:2013,astropy:2018},
\href{https://numpy.org/}{\texttt{numpy}} \citep{harris2020array},
\href{https://www.scipy.org/index.html}{\texttt{scipy}} \citep{2020SciPy-NMeth},
\href{https://matplotlib.org/stable/index.html}{\texttt{matplotlib}} \citep{matplotlib},
\href{https://spectres.readthedocs.io/en/latest/}{\texttt{spectres}} \citep{spectres}
}


\bibliography{sample63}{}
\bibliographystyle{aasjournal}

\end{document}